\documentclass[aps,pra,reprint,nofootinbib,unsortedaddress]{revtex4-2}

\usepackage{amsmath}
\usepackage{mathtools,cases}
\usepackage{epsfig}
\usepackage{amssymb}
\usepackage{graphicx}% Include figure files
\usepackage{dcolumn}% Align table columns on decimal point
\usepackage{bm}% bold math
\usepackage{lipsum}
\usepackage{upgreek}
\begin{document}

%%\unnumbered% uncomment this for unnumbered level heads
%Dynamic dielectric metasurfaces based on lattice resonances: tuning and switching effects via water
% Enabling optical tunability 
% Optical tunability and dynamics of Mie-type surface lattice resonances in a non-homogeneous environment.
\title{Dynamic dielectric metasurfaces via control of surface lattice resonances in non-homogeneous environment}

\author{Izzatjon Allayarov$^{\dag1,2,3}$, Andrey B. Evlyukhin$^{\dag3,4*}$, Diane J. Roth$^{\dag5}$, Boris Chichkov$^{3,4}$, Anatoly V. Zayats$^{5*}$ and Antonio Cal{\`a} Lesina$^{1,2,3*}$} 

\affiliation{$^1$Hannover Centre for Optical Technologies, Leibniz University Hannover, Nienburger Str. 17, D-30167 Hannover, Germany}
\affiliation{$^2$Institute of Transport and Automation Technology, Leibniz University Hannover, An der Universit{\"a}t 2, D-30823 Garbsen, Germany}
\affiliation{$^3$Cluster of Excellence PhoenixD, Leibniz University Hannover, Welfengarten 1A, D-30167 Hannover, Germany}
\affiliation{$^4$Institute of Quantum Optics, Leibniz University Hannover, Welfengarten 1, D-30167 Hannover, Germany}
\affiliation{$^5$Department of Physics and London Centre for Nanotechnology, King's College London, Strand, WC2R 2LS London, UK}

\begin{abstract}
	Dynamic control of metamaterials and metasurfaces is crucial for many photonic technologies, such as flat lenses, displays, augmented reality devices, and beam steering, to name a few. The dynamic response is typically achieved by controlling the phase and/or amplitude of individual meta-atom resonances using electro-optic, phase-change or nonlinear effects. Here, we propose and demonstrate a new practical strategy for the dynamic control of the resonant interaction of light with dielectric metasurfaces, exploiting the dependence of the interaction between meta-atoms in the array on the inhomogeneity of the surrounding medium. The revealed tuning mechanisms are based on the concept of the surface lattice resonance (SLR), the development of which strongly depends on the difference between permittivities of superstrate and  substrate materials. We experimentally demonstrate surface lattice resonances in dielectric (Si) metasurfaces, and reveal two tuning mechanisms corresponding to shifting or damping of the SLR in optofluidic environment. The demonstrated dynamic tuning effect with the observed vivid colour changes may provide a dynamic metasurface approach with high spectral selectivity and enhanced sensitivity for sensors, as well as high-resolution for small pixel size displays. %, as it is governed by short-range SLRs involving only a few meta-atoms. 

%The strategy exploits the dependence of the surface lattice resonance (SLR) on the homogeneity of the surrounding medium. The revealed tuning mechanisms are theoretically explained via the concept of the dipole lattice sum (DLS) in the case of substrate presence. We demonstrate that the value of the DLS, which determines the conditions of SLR excitation and its spectral position, strongly depends on the difference between dielectric constants of superstrate and  substrate materials. By using a time-dependent change in this difference, dynamic control of the excitation and spectral tuning of SLR can be realized. As a practical and demonstrative example, we investigate the optical response of high-index silicon metasurfaces as a function of progressive immersion in a liquid such as water. Considering two superstrates, such as air and water, we have two tuning options: the SLR exists for both superstrates or only for one, corresponding to shifting and switching of the SLR. Metasurface designs are fabricated and experimentally validated. Both tuning mechanisms are demonstrated considering immersion in water (static response) and evaporation of water (dynamic response). In the latter case, as water transitions through the metasurface, a spectrum variation is recorded, proving the optical dynamic response.  The results obtained  form the basis for a novel optical tuning approach that leads to metasurfaces with high spectral selectivity, enhanced sensitivity, and short-range SLRs involving only a few nanostructures for higher-resolution displays.

\end{abstract}

%\keywords{keyword1, Keyword2, Keyword3, Keyword4}

\maketitle
\def\thefootnote{*}\footnotetext{Corresponding author(s). E-mail(s): evlyukhin@iqo.uni-hannover.de, anatoly.zayats@kcl.ac.uk, antonio.calalesina@hot.uni-hannover.de}\def\thefootnote{\arabic{footnote}}
\def\thefootnote{$\dagger$}\footnotetext{These authors contributed equally to this work}\def\thefootnote{\arabic{footnote}}
	
The ultimate goal of nanophotonic metasurfaces is to achieve a dynamically tuneable optical response required to enable applications in beam steering, adaptive lenses, augmented reality, optical switches, and displays \cite{Shaltout2019a}. Tuneability of metasurfaces has been demonstrated using carrier accumulation/depletion via voltage gating, liquid crystals, mechanical actuation, chemical reactions, phase change materials, nonlinear effects in time-varying materials, liquid injection through microfluidic channels \cite{Li2022c,Liu2019,Sun2018}, and immersion in liquid \cite{Li2022b,Wan2022}. In all these cases, primarily the optical resonances of the individual plasmonic or dielectric meta-atoms are affected, influencing the amplitude, phase or direction of the transmitted or reflected light.

Collective effects in periodic 2D arrays of nanoparticles, such as surface lattice resonances (SLRs), have been investigated to achieve optical response typically not obtainable with single particles, such as resonances with high quality factor \cite{Bin-Alam2021a}. An SLR occurs if the first diffraction order of a periodic array becomes an evanescent wave propagating in the plane of the array, the so-called Rayleigh anomaly (RA).
%, which practically happens when the excitation wavelength in the medium matches the period of the array.  --AZ: but it depends on angle of incidence, so I removed the statement.
The sensitivity of the SLR to the refractive index of the environment and, as a consequence, the modification of the SLR spectral position and its quality factor, makes this collective effect very attractive in sensing applications \cite{kuznetsov2011laser,offermans2011universal,thackray2014narrow,aristov2014laser,Kravets2018a}.

The SLR concept was initially developed for arrays of metallic nanoparticles with electric dipolar response \cite{zou2004,markel2005divergence,auguie2008collective,offermans2011universal}, and later extended to metallic particles supporting high-order plasmonic multipoles \cite{giannini2010lighting,Evlyukhin2012}. 
Design strategies to achieve multi-resonant plasmonic metasurfaces based on SLRs were also reported \cite{Reshef2019,Reshef2022}. Tuneability of SLRs was achieved via post-fabrication thermal control \cite{Kelavuori2022} as well as using electrically controlled liquid crystals \cite{Sharma2022} in plasmonic metasurfaces. The concept of the lattice resonances has also been theoretically expanded to dielectric metasurfaces supporting Mie resonances \cite{Evlyukhin2010,Castellanos2019,Babicheva2021}. 

%Despite the general understanding that SLRs are sensitive to the external environment and tend to degrade as the surrounding medium becomes inhomogeneous, an approach for optical tunability based on this concept is elusive. For this reason, in most of the above-mentioned theoretical works a superstrate that matches the refractive index of the substrate was chosen. In some cases, the presence of the substrate was even ignored, and the approximation of a homogeneous environment was made for ease of calculation. In general, the presence of a substrate can significantly change the optical properties of a metasurface compared to the case of a homogeneous environment. This is especially true for SLRs, which can be completely suppressed in systems with an asymmetric environment due to restricted long-range interactions in the array \cite{mahi2017depth,Auguie2010}. Here, we turn the presence of the substrate to our advantage, we point out that the SLR is not immediately suppressed as the surrounding becomes non-homogeneous, and we exploit the dependence of the SLR excitation on the superstrate/substrate dielectric contrast to enable optical tunability and dynamical optical control. 

In most of the above-mentioned works on SLRs, the approximation of a homogeneous environment was used for ease of simulations. However, the presence of a substrate can significantly modify the optical properties of a metasurface. This is especially true for SLRs, which can be completely suppressed when the metasurface is in a non-homogeneous medium \cite{mahi2017depth,Auguie2010}. Nevertheless, the presence of the substrate can be an advantage, providing opportunities to control the SLR excitation with the superstrate/substrate dielectric contrast in order to enable tuneability and dynamic control of the optical response. 

In this paper, we describe and demonstrate this dynamic tuning principle on the example of a silicon metasurface on a glass substrate, and introduce an optofluidic platform using a variable water level as a strategy to control the refractive index of the superstrate. Although SLRs in dielectric metasurfaces have not been reported experimentally until now, the choice of dielectric metasurfaces is motivated by negligible material losses compared to plasmonic metasurfaces and the possible future extensions of the approach to other Mie-type multipoles. The use of water makes the tuning mechanism reversible because, contrary to oil, water can be totally removed from the metasurface. Liquid environments provide flexibility in using microfluidic control~\cite{Li2022c}, electrowetting effect~\cite{Wu2019} or simply through thermal evaporation/condensation, and can be extended to a variety of liquids beyond water. Furthermore, the switching speed of the metasurface can be regulated by adopting an appropriate microfluidic control technique.

Numerical, theoretical and experimental investigations show that there are two mechanisms to control the optical response of dielectric metasurfaces via SLRs: one providing suppression of the resonance transmission/reflection and another resulting in the spectral shift of the resonance. Both revealed mechanisms provide vivid changes in the colouring of the metasurfaces with the change of the water environment.
%how the variable level of water on top of the metasurface enables optical switching and shifting in reflection and transmission. We start by validating these ideas via numerical simulations. Then a theoretical model is presented to explain the details behind the emergence of such tuning mechanisms. Finally, the metasurface designs where these scenarios are more prominent were fabricated and experimentally validated.
The results demonstrate new practical possibilities for tuning light-matter interaction at the subwavelength scale, which can find application in dynamic generation of colours for ultra-high resolution displays, high sensitivity refractive index sensors and anticounterfeiting protection. %techniques for nanometric sensing. 

\section*{Results}\label{results}
\subsection*{Numerical simulations}\label{meta_num}
\begin{figure*}[!ht]
  \includegraphics[width=15cm]{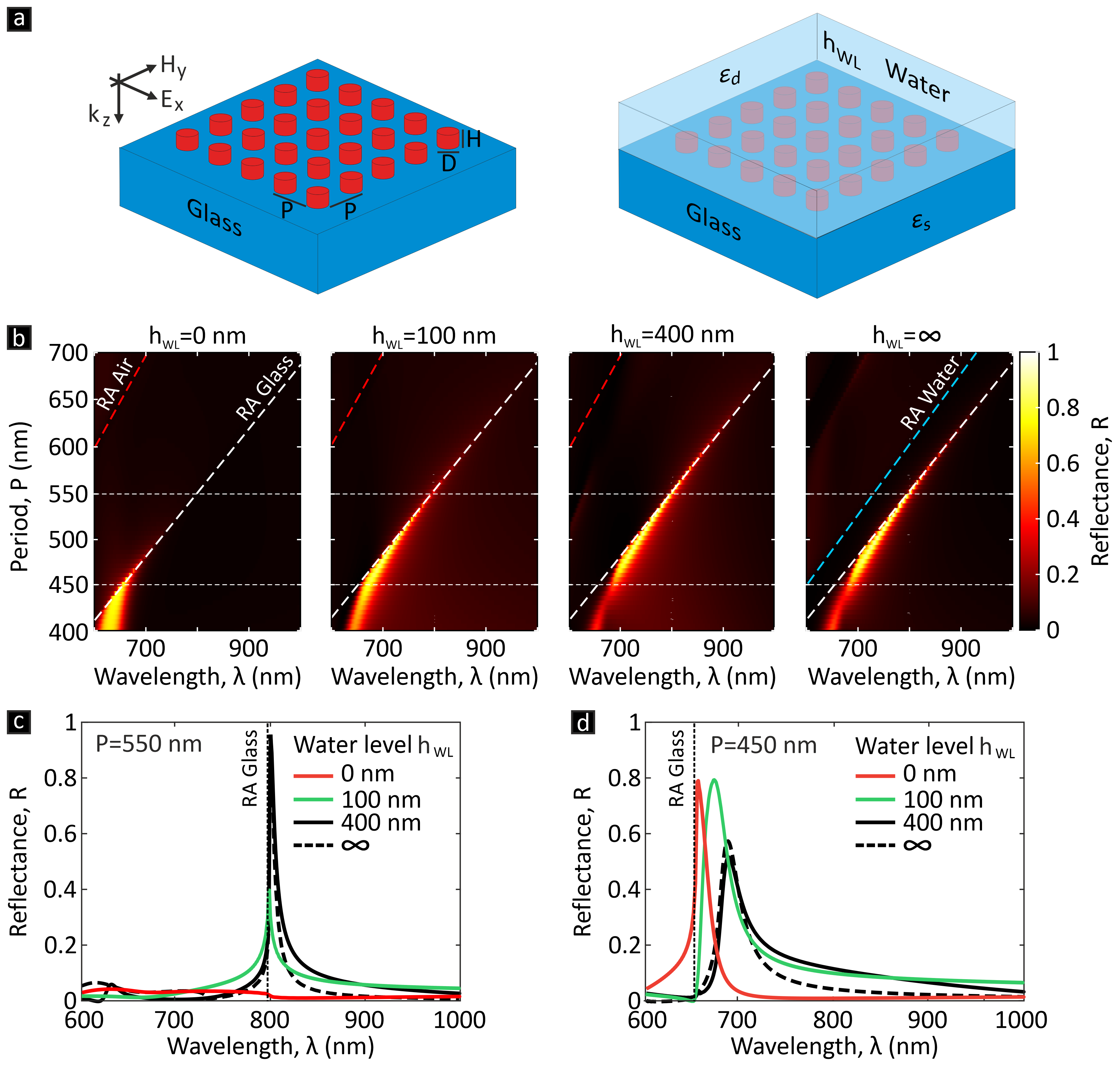}
  \caption{\textbf{Optical response of a metasurface with a dynamic superstrate.} (a) Schematic of a metasurface: Si nanodisks of diameter $D$ and height $H$ are arranged in a square array on a glass substrate with periodicity $P$ along the $x$ and $y$ directions. The optical response of the system is controlled by the level of water $h_{\mathrm{WL}}$ on top of a metasurface. 
  %If not specified explicitly, we assume the nanodisks height $H$=100~nm. 
  The metasurface is illuminated at normal incidence, with $x$-linearly polarized light and reflection or transmission is observed. (b) Dependence of reflectance spectra of an infinite array of Si nanodisks with $D$=200~nm and $H$=100~nm on a period $P$ for different water levels $h_{\mathrm{WL}}$. The red (blue) and white dashed lines display the RA conditions for the air (water) superstrate and glass substrate, respectively.
  (c,d) Reflectance spectra of the metasurfaces with (c) $P$=550~nm and (d) $P$=450~nm, cross-sections of Fig.~\ref{fig:idea}b (white horizontal dashed lines in Fig.~\ref{fig:idea}b), illustrating two mechanisms for switching at the limiting values of $h_{\mathrm{WL}}$: suppression of resonance in (c) and a spectral shift of resonance in (d).}
  \label{fig:idea}
\end{figure*}

We consider an infinite metasurface composed of Si nanodisks of diameter $D$ and height $H$ placed on a semi-infinite glass substrate with a square lattice of period $P$, and illuminated by a linearly polarized plane wave at normal incidence (Fig.~\ref{fig:idea}a). 
The simulated reflectance of the metasurface strongly depends on both the period and the water level $h_{\mathrm{WL}}$ above the substrate (Fig.~\ref{fig:idea}(b)). The strongest reflectance is observed near the long wavelength side of the RA related to the glass substrate (RA glass), when the SLR appears. In fact, SLRs can only exist in the diffractionless regime, and we will refer to the spectral range on the right of the substrate RA as the SLR region.  
%For convenience, we have indicated in Fig.~\ref{fig:idea}b the RAs of the main diffraction order in the superstrate $\lambda^{\rm RA}_{\rm d}$ (red/blue dashed lines) and substrate $\lambda^{\rm RA}_{\rm s}$ (white dashed lines), which divide the spectral region into three subregions: i) $\lambda <\lambda^{\rm RA}_{\rm d}$, ii) $\lambda^{\rm RA}_{\rm d}<\lambda<\lambda^{\rm RA}_{\rm s}$ and iii) $\lambda>\lambda^{\rm RA}_{\rm s}$. 
At the same time, in the range on the left of the substrate RA, SLRs are absent due to the diffraction scattering into the superstrate and/or substrate~\cite{Auguie2010,mahi2017depth}.  

Depending on the period of the metasurface, one can distinguish two scenarios for the development of a resonant response, as highlighted by the two horizontal dashed lines in Fig.~\ref{fig:idea}b, which are separately plotted in Figs.~\ref{fig:idea}c,d. For longer periods, e.g., $P$=550~nm (Fig.~\ref{fig:idea}c), there is no resonance when the metasurface is in air ($h_{\mathrm{WL}}=0$). The sharp resonance starts emerging only by adding water above a certain threshold level, approximately $h_{\mathrm{WL}}\geq H=100$~nm, i.e., disks are fully surrounded by water. The quality factor and spectral position of the resonance are indicative of an SLR excitation. For shorter periods, e.g., $P$=450~nm (Fig.~\ref{fig:idea}d), the resonance already exists with air as a superstrate and increasing the water level leads to a red-shift of the resonance. In both cases, increasing the water level above $h_{\mathrm{WL}}\geq 400$~nm does not change significantly the shape and position of the resonances, and the superstrate can be assumed as a homogeneous water environment ($h_{\mathrm{WL}} = \infty$). 
%Indeed, one can apply an effective medium approach and replace air and water layers with a single homogeneous environment~\cite{Qiu2002}. 
Although a finite water level may introduce Fabry–P{\'e}rot resonances~\cite{Reshef2019}, they would appear on the left side of the substrate RA, and become significant only when the water level is comparable with the wavelength. This is not a regime under investigation in this work. In the next section, based on an analytical model, we explain how these two different regimes of the dynamic response emerge from the SLR excitation and evolve with the refractive index of the superstrate.

\subsection*{Analytical model}\label{meta_theor}

Silicon disks forming a metasurface (Fig.~\ref{fig:idea}a) exhibit mainly electric and magnetic dipole responses in the considered spectral range, regardless of the illumination and surroundings (Figs.~S1--S5 in Supplementary Information). Therefore, the coupled-dipole model~\cite{mulholland1994light,merchiers2007light,bendana2009confined} can be used for the analytical investigation of the emergence of the metasurface collective resonances observed numerically in Fig.~\ref{fig:idea}b. Since the electric dipole coupling dominates over the magnetic one for the metasurface in a homogeneous surrounding (see Section 2 of Supplementary Information), and the presence of a glass substrate provides only a very weak coupling between electric and magnetic dipoles \cite{miroshnichenko2015substrate}, we will consider only coupling between electric dipoles of the Si disks.

Due to the in-plane translation symmetry and the normal illumination condition (Fig.~\ref{fig:idea}a), every particle of an infinite metasurface will have the same electric dipole moment ${\bf p}=(p_x,0,0)$ defined as \cite{Abajo2007,Evlyukhin2010}
\begin{equation}\label{ed2_s}
    p_x=\frac{\varepsilon_0\varepsilon_{\rm d}\tilde E_x}{1/\tilde\alpha_{xx}^{\rm p}-\tilde S} \, ,
\end{equation}
where $\varepsilon_0$, $\varepsilon_{\rm d}$ and $\varepsilon_{\rm s}$ are the vacuum permittivity, and superstrate and substrate relative permittivities, respectively, $\tilde\alpha_{xx}^{\rm p}$ is the electric dipole polarizability of a single disk near a substrate, $\tilde E_x$ is the incident electric field at the geometrical disk centre (in the absence of the disk and in the presence of the substrate), and $\tilde S$ is the dipole lattice sum in the presence of a substrate and associated with the electromagnetic interaction (coupling) between dipolar particles of the metasurface.

This dipolar system is in resonance when the denominator of Eq.~\eqref{ed2_s} vanishes.
In the presence of damping associated with material and radiative losses, the conditions for the SLR excitation are
\begin{subequations}\label{re_s}
\begin{numcases}{}
    \mathrm{Re}(1/\tilde \alpha_{xx}^{\rm p})=\mathrm{Re}(\tilde S), \label{re_s:re} \\
    \mathrm{min}\{\lvert\mathrm{Im}(1/\tilde\alpha_{xx}^{\rm p}-\tilde S)\rvert\}. \label{re_s:im}
\end{numcases}
\end{subequations}
These conditions include the properties of individual particles via their polarizability and the properties of the array via the dipole lattice sum, which does not depend on the particle properties \cite{Kravets2018a}. The conditions in Eq.~\eqref{re_s} can only be satisfied in the SLR region ($\lambda>\lambda^{\rm RA}_{\rm s}$).
A moderate (air/glass or water/glass) dielectric contrast between a substrate and a superstrate $\Delta\varepsilon=(\varepsilon_{\rm s}-\varepsilon_{\rm d})$ does not significantly affect a broad resonance of $\tilde\alpha_{xx}^{\rm p}$, causing only its red-shift (Fig.~S5). Therefore, the appearance of narrow resonances in the optical response of a metasurface with the changes of a water level (Fig.~\ref{fig:idea}b) are related to the effect of $\Delta\varepsilon$ on the lattice sum $\tilde S$ (i.e., on the inter-particle interaction).

\begin{figure*}[!ht]
	\includegraphics[width=15cm]{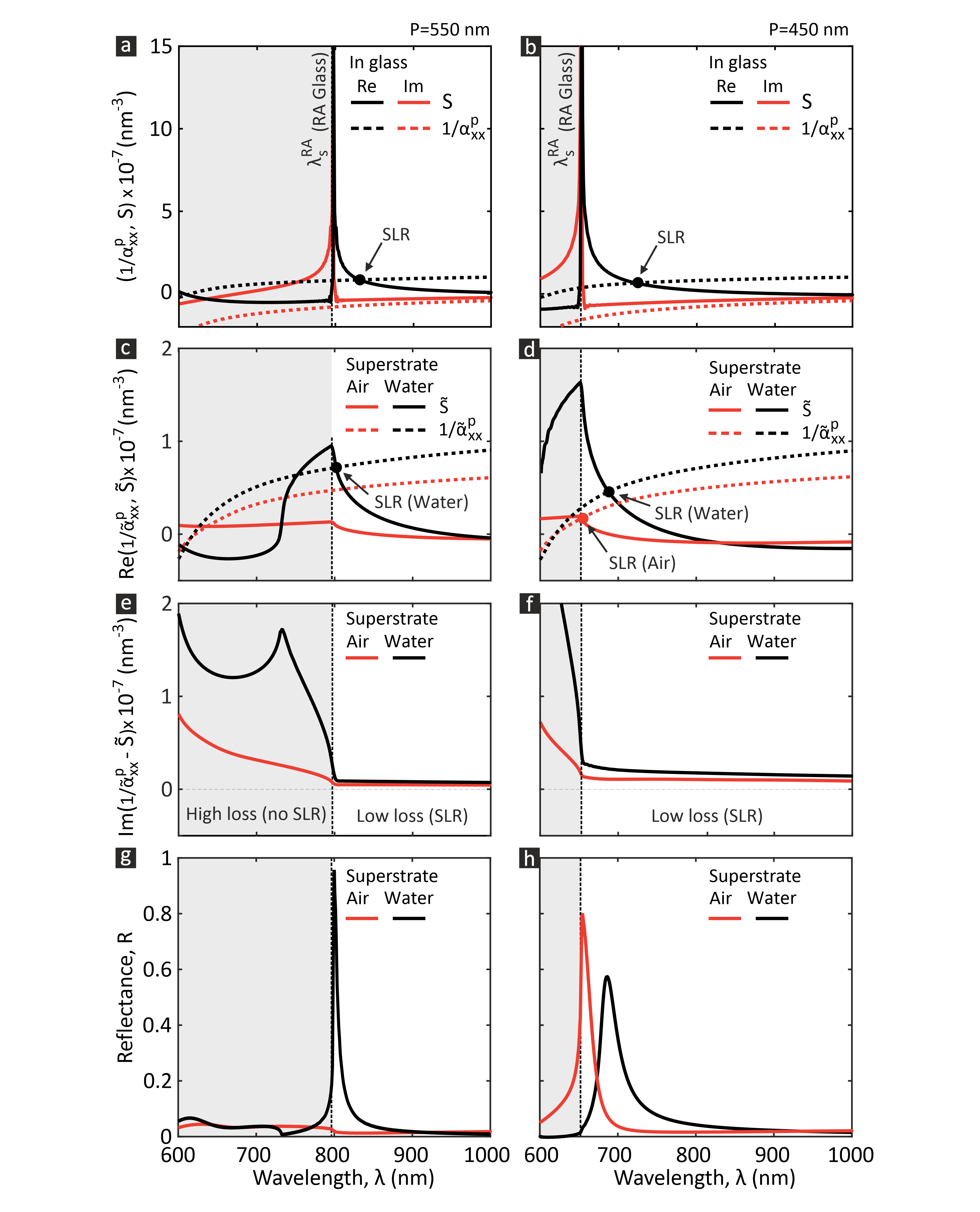}
	\caption{\textbf{Resonant condition analysis.} (a,b) Spectra of the inverse electric dipole polarizability $1/\alpha_{xx}^{\rm p}$ and the full lattice sum $S$ for a metasurface in a homogeneous glass environment: (black) real and (red) imaginary parts. (c,d) Spectra of $1/\tilde \alpha_{xx}^{\rm p}$ and $\tilde S$ for a metasurface on a glass substrate with (red) air or (black) water superstrate (only the real part is shown). (e,f) Spectra of the difference of $1/\tilde \alpha_{xx}^{\rm p}$ and $\tilde S$ for a metasurface in the case of (red) air/glass and (black) water/glass superstrate/substrate combinations (only the imaginary part is shown). (g,h) Reflectance spectra of a metasurface for (red) air/glass and (black) water/glass cases. (a,c,e,g) correspond to $P$=550~nm and (b,d,f,h) correspond to $P$=450~nm; in all panels $D$=200~nm and $H$=100~nm; the superstrate and substrate are considered to be semi-infinite. The vertical black dashed line $\lambda_{\rm s}^{\rm RA}$ indicates the position of the RA in the glass substrate. Filled circles in (a,b,c,d) indicate the spectral position of SLRs.}
	\label{fig:rc_sub}
\end{figure*} 

For a homogeneous glass environment ($\Delta\varepsilon=0$), the lattice sum $S$ has a divergent singularity at the RA (Figs.~\ref{fig:rc_sub}a,b), due to radiative (far-field) coupling between nanoparticles in the array (Fig.~S6). Therefore, the resonant conditions in Eq.~\eqref{re_s} are satisfied and the SLR is excited, as graphically illustrated by the intersection between $1/\alpha_{xx}^{\rm p}$ and $S$ in the SLR region in Figs.~\ref{fig:rc_sub}a,b.  

In the presence of a substrate, the lattice sum $\tilde S$ becomes finite and depends on the dielectric contrast $\Delta\varepsilon$ and the array periodicity. In this case, in addition to $S$, the lattice sum also includes a contribution that takes into account the reflection from the substrate $S^{\rm R}$, so that $\tilde S=S+S^{\rm R}$.

%Must be fixed
%Contrary to the lattice sum $S$ in a homogeneous medium, which can be calculated analytically, the lattice sum in the presence of a substrate $\tilde S$ must be calculated numerically, and only its far-field contribution can be derived analytically. 
%While the far-field contribution in $S$ diverges (Fig.~S6), the far-field term in $\tilde S$, i.e., $\tilde S^{\rm FF}$ does not diverge which implies the non divergence of the total lattice sum (Fig. S9).
%Since the possible divergence of S can come from FF contribution, we provide an analytical formula (SI), Eq . 3 which demonstrates this point due to reflection from the substrate, i.e., the $r_l$ term.

For both $S$ and $\tilde S$, the divergence of the lattice sum only depends on the far-field contribution. The part $\tilde S^{\rm FF}$ corresponding to the far-field inter-particle coupling in the metasurface on a substrate can be evaluated as (see Section~3 of Supplementary Information)
\begin{align}\label{s_app}
    \tilde S^{\rm FF} \approx \sum_{l\neq0}^\infty S^{\rm FF}_l \Bigl (1 + r_l^{({\rm s})} \Bigr),
\end{align}
where $S^{\rm FF}_l$ is the $l$-th far-field lattice term in the homogeneous environment (superstrate) and $r_l^{({\rm s})}$ is the reflection coefficient for the s-polarized wave generated by the electric dipole located at a lattice $l$-th node and propagating to a node with $l=0$ (the distance between these points is $R_l$). The reflection coefficient is then
\begin{equation}\label{rsl}
    r^{({\rm s})}_l=\frac{1-\sqrt{R_l^2/\Tilde{R}^2+1}}{1+\sqrt{R_l^2/\Tilde{R}^2+1}},
\end{equation}
where $\Tilde{R}=z_{\rm p}\sqrt{{\varepsilon_{\rm d}}/{\lvert\Delta\varepsilon\rvert}}$ is the effective distance determined by the dielectric contrast $\Delta\varepsilon$ with $z_{\rm p}$ being the distance from a nanodisk centre to the substrate. The value of $\Tilde{R}$ can be used to estimate the role of the long-range (far-field) interaction between nanoparticles in metasurfaces located on a substrate. Importantly, for the  metasurfaces considered in  Fig.~\ref{fig:idea}, $\Tilde{R}$ is much smaller than the metasurface periods, i.e., $R_l>>\tilde{R}$. Therefore, from Eq.~(\ref{rsl}) one obtains that the reflection coefficient $r_l^{({\rm s})}\simeq-1$ and the far-field sum Eq.~\eqref{s_app} will be determined by only a finite number of terms due to the suppression of the far-field coupling between the nanoparticles. 
As a result, the lattice sum in the presence of a substrate does not diverge, having a finite value at the substrate RA \cite{Auguie2010}. However, as shown in (Figs.~\ref{fig:rc_sub}c,d), the maximum value of the real part of the lattice sum is still obtained at the substrate RA ($\lambda_{\rm s}^{\rm RA}$) and decreases with the increase of the dielectric contrast $\Delta\varepsilon$.

%The decrease of the lattice sum with increasing $\Delta\varepsilon$ means that the number of particles participating in the effective inter-particle coupling decreases due to suppression of long-range interaction, which is the origin of what we call short-range SLRs.
%Controlling the number of interacting particles via $\Delta\varepsilon$ enables the tuning mechanisms described in this paper.
On one hand, a decreasing lattice sum can lead to $\mathrm{Re}(\tilde S)<\mathrm{Re}(1/\tilde \alpha_{xx}^{\rm p})$ at the substrate RA, where the conditions in Eq.~\eqref{re_s} can not be satisfied, and the SLR is not excited. On the other hand, if the lattice sum is large enough to obtain $\mathrm{Re}(\tilde S)>\mathrm{Re}(1/\tilde \alpha_{xx}^{\rm p})$ at the substrate RA, then the conditions in Eq.~\eqref{re_s} are satisfied and the lattice resonance can exist in the SLR region. %Notice that a finite value of $\tilde S$ is sufficient to obtain an SLR. 
This is due to the fact that $\mathrm{Re}(\tilde S)$ monotonically decreases and $\mathrm{Re}(1/\tilde \alpha_{xx}^{\rm p})$ monotonically increases with the wavelength in the SLR region. 
Such behaviour is observed in Fig.~\ref{fig:rc_sub}c, where the condition Eq.~\eqref{re_s}a is satisfied only for the metasurface with a small dielectric contrast (water/glass) and period $P$=550~nm. The second condition Eq.~\eqref{re_s}b is illustrated in  Fig.~\ref{fig:rc_sub}e and is satisfied in the SLR region for both superstrates. %: for the air/glass configuration, the SLR conditions are not satisfied in the SLR region, whereas they are satisfied for the water/glass configuration. 
As a result, the reflection spectrum for the water superstrate has a narrow resonant peak with a theoretical quality factor of over 100 (Fig.~\ref{fig:rc_sub}g) corresponding to the full circle in Fig.~\ref{fig:rc_sub}c. We note that the quality factor of the system can be increased by an order of magnitude by further optimization of the structure (e.g., period, superstrate material). At the same time, outside of the SLR region, the SLR conditions are not fully satisfied; the condition Eq.~\eqref{re_s}a is fulfilled (Fig.~\ref{fig:rc_sub}c) while the condition Eq.~\eqref{re_s}b is not met due to high losses (Fig.~\ref{fig:rc_sub}e). %This mechanism corresponds to on/off switching. % is not (empty circles).

For the metasurface with a smaller period $P$=450~nm, the lattice sum is larger than the inverse polarizability at the substrate RA for both superstrates. As a result, the conditions for the SLRs are satisfied in the SLR region for both air/glass and water/glass configurations (Figs.~\ref{fig:rc_sub}d,f) and the SLRs are excited with a spectral shift of the resonance depending on the dielectric contrast $\Delta\varepsilon$ (Fig.~\ref{fig:rc_sub}h). %This mechanism corresponds to colour tuning.

%Having considered the numerical and analytical models, the following general conclusions can be derived. In the case of a metasurface located in an inhomogeneous environment (superstrate/substrate configuration), the effective number of nanoparticles participating in the simultaneous interaction depends on the value of the dielectric contrast between substrate and superstrate $\Delta\varepsilon$ and decreases when the contrast increases. Such dependence leads to the suppression of SLRs for the configurations with sufficiently large $\Delta\varepsilon$. As a result, the collective resonant response of a metasurface with the change of $\Delta\varepsilon$ can evolve according to two scenarios.  

For the practical implementation of these two scenarios and their use for dynamic control of metasurface response, the following strategy can be followed.
First of all, we need that for a given $\Delta\varepsilon$, ${\rm Re}(\tilde S)>{\rm Re}(1/\tilde\alpha_{xx}^{\rm p})$ at the substrate RA, so that the SLR is excited in the SLR region. Second we can have: 
\noindent (i) On/off switching. If we increase $\Delta\varepsilon$ enough to reach the condition ${\rm Re}(\tilde S)<{\rm Re}(1/\tilde\alpha_{xx}^{\rm p})$ at the substrate RA, then the SLR is switched off. 
\noindent (ii) Resonance shift. If we decrease $\Delta\varepsilon$, the SLR remains excited but with a red shift.
These switching and shifting effects, which may result in colour tuning, are shown in Fig.~\ref{fig:rc_sub}g and Fig.~\ref{fig:rc_sub}h, respectively, for changing the superstrate from air to water.

\section*{Experimental demonstration}
\subsection*{Static optical response}
\begin{figure*}[!ht]
\begin{center}
  \includegraphics[width=15cm]{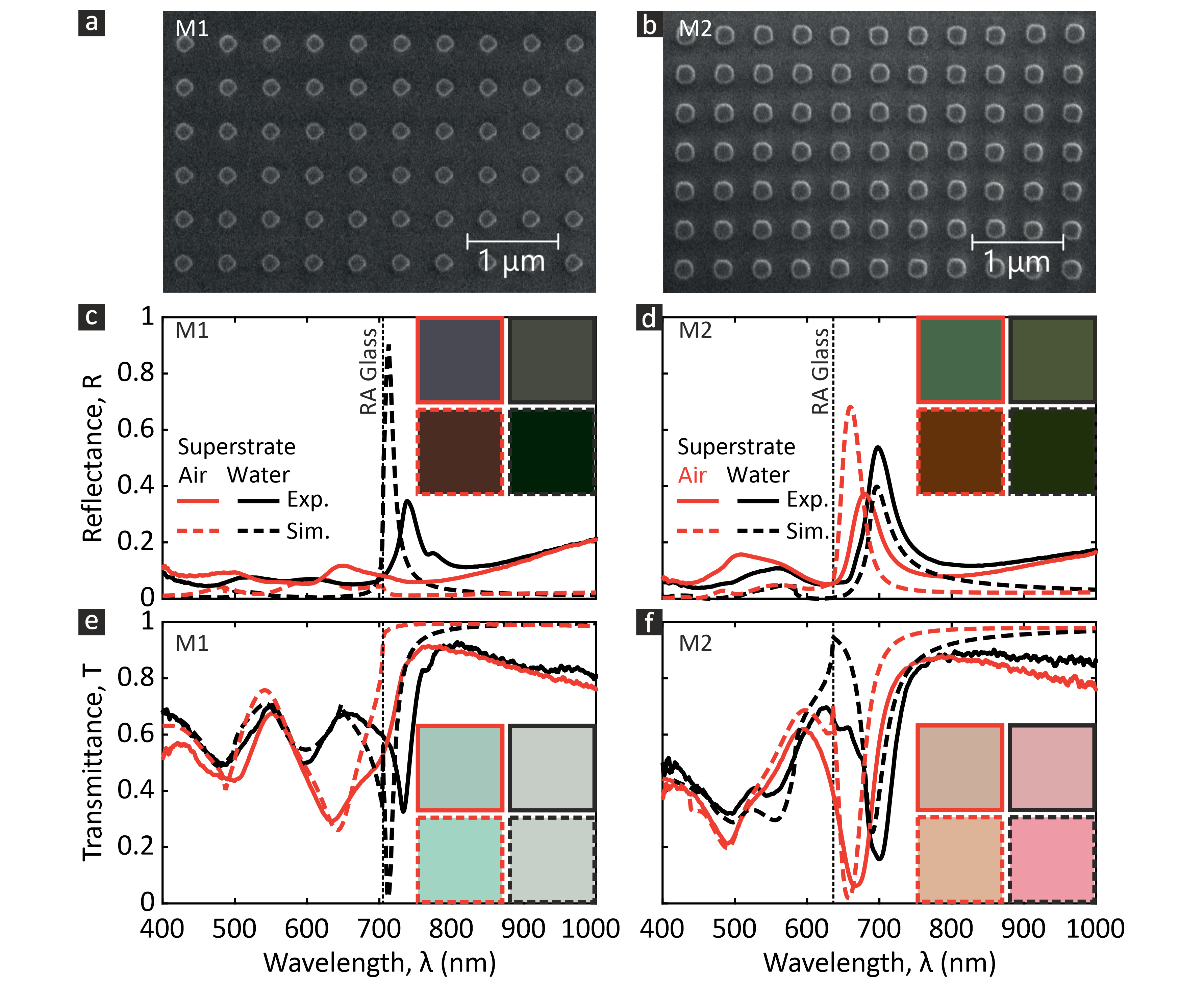}
  \caption{\textbf{Transmission and reflection of the metasurface with different superstrates.} (a,b) Scanning electron microscope (SEM) images of the fabricated metasurfaces: (a) M1  with $D$=185~nm and $P$=490~nm and (b) M2 with $D$=220~nm and $P$=440~nm. (c,d) Reflectance and (e,f) transmittance of metasurfaces (c,e) M1 and (d,f) M2 for (red) air and (black) water superstrates: (solid lines) experiment, (dashed lines) simulations. The vertical dashed lines correspond to the spectral position of the RAs of a glass substrate $\lambda_{\rm s}^{\rm RA}$. Insets show the colour representation of the corresponding spectra.}
  \label{fig:steadyComp}
  \end{center}
\end{figure*}
To experimentally illustrate the switching and shifting scenarios of lattice resonances, metasurfaces consisting of $H$=100~nm thick polycrystalline silicon nanodisks on a glass substrate were fabricated using electron beam lithography (see Methods for the details of fabrication and geometrical parameters). Several diameters and periods of the nanodisks (see Table~\ref{tab:tab1} in Methods with a selection of four metasurfaces, M1-M4) were considered in order to investigate tuneability in both switching and shifting scenarios. 
%The steady-state response of the lattice resonances in the metasurfaces is demonstrated by measuring reflectance and transmittance spectra of the samples in air and immersed in water. For both measurements, the light is normally incident on the sample, polarised along the direction of periodicity of the array, and weakly focused onto the sample. The transmitted or reflected light is then coupled to a spectrometer via an optical fibre. In both cases, a small portion of the collected light is also diverted onto a CCD camera to record images or movies of the sample. See the Methods section for more details about the experimental procedures.
Transmission and reflection spectra as well as CCD images of the metasurface were recorded to demonstrate the dependence of the optical properties on the superstrate refractive index (see Methods for the details of the experimental set-up).

For the metasurface M1 (Fig.~\ref{fig:steadyComp}a), the reflectance spectrum of the metasurface in air does not feature any resonances since lattice resonance is absent due to the large air/glass dielectric contrast (Fig.~\ref{fig:steadyComp}c). Immersing the metasurface in water, therefore reducing the dielectric contrast (water/glass system), leads to the strong switching of the lattice resonance around a wavelength of 730~nm and a strong reflection peak. Similar switching behaviour is observed in transmission (Fig.~\ref{fig:steadyComp}e); note that the transmission spectra show also additional resonances due to the absorption of silicon at shorter wavelengths. Measured spectra are in overall good agreement with the modelling. Simulated transmittance and reflectance spectra were also converted to their colour appearance by using CIE 1931 XYZ colour space~\cite{Smith1931} and the simulated visual appearance can be directly compared to the experimental colours (see insets in Figs.~\ref{fig:steadyComp}c,d,e,f). While absorption in silicon in the visible range prevents us from obtaining a rich gamut of colours, this can be improved by using other materials, such as for example silicon nitride. The smaller magnitude of the resonances observed in the experiments as well as the slight shift of the resonances compared to the model spectra could be due to the non-perfect nature of the fabrication process, slight variations in the geometrical parameters of the nanodisks, and the finite size of the fabricated metasurface. Note that $R+T<1$ in Fig.~\ref{fig:steadyComp} since in the experimental measurements the collection angle is $<180^{\circ}$. In simulations, the fraction of reflectance or transmittance within this collection angle is considered for fair comparison.

The second type of optical response tuning is demonstrated with a metasurface M2 (Fig.~\ref{fig:steadyComp}b), for which the change in surrounding refractive index between air and water leads to a red shift of the lattice resonance from $\lambda$=680~nm in air to $\lambda$=700~nm in water (Figs.~\ref{fig:steadyComp}b,d). Again, a very good agreement between experimental and theoretical results is observed. The equivalent of Fig.~\ref{fig:steadyComp} with the static optical response of the metasurfaces M3 and M4 is reported in Fig. S10.

\subsection*{Dynamic tuneability}
\begin{figure*}[!ht]
    \includegraphics[width=15cm]{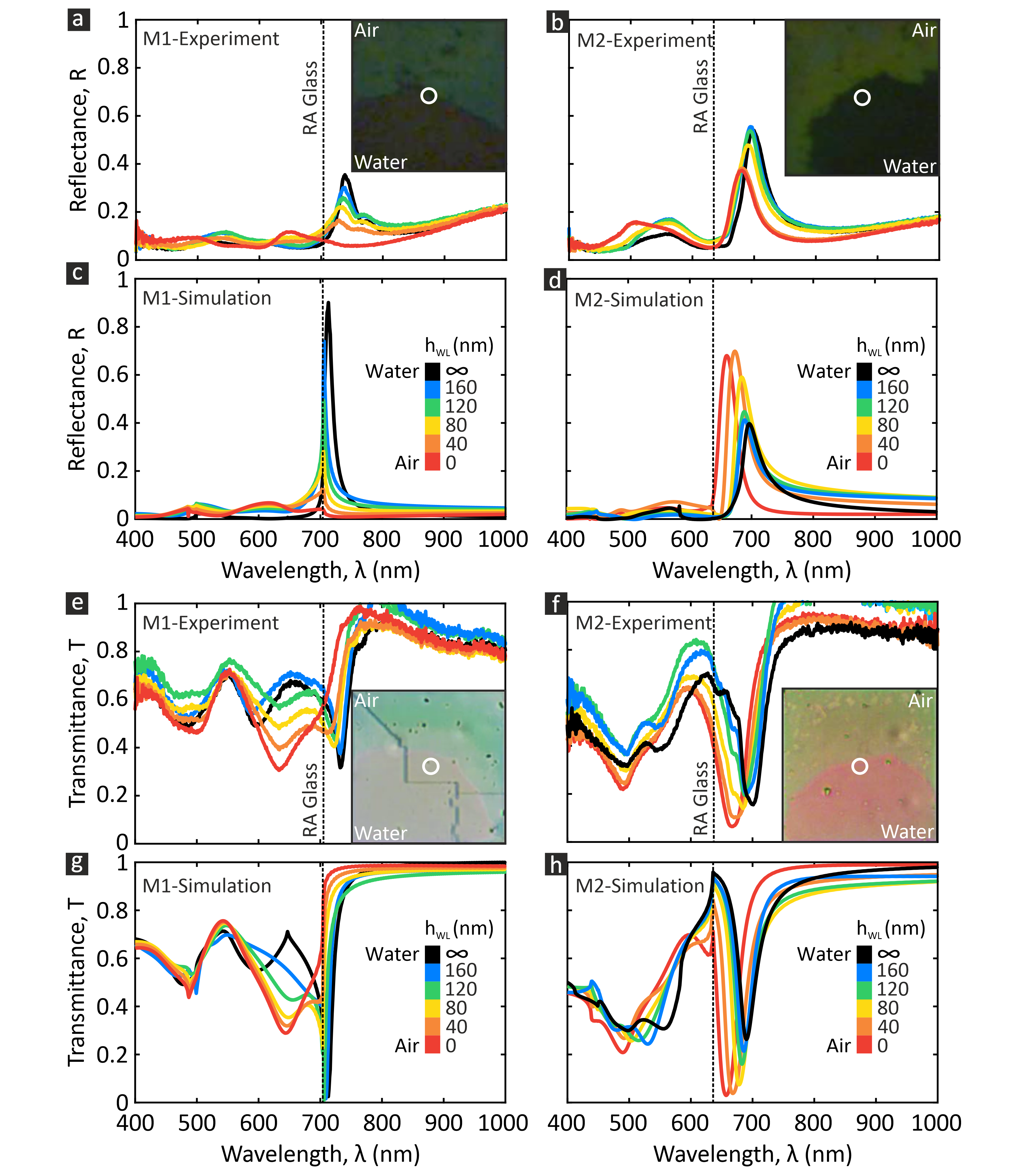}
  \caption{\textbf{Dynamic tuning of the transmission and reflection of the metasurface via superstrate control.} (a--d) Reflectance and (e--h) transmittance of metasurfaces (a,c,e,g) M1 and (b,d,f,h) M2 while dynamically changing the superstrate between air and water: (a,b,e,f) experimental spectra measured every 100~ms (snapshot of the Supplementary Movies M1R, M2R, M1T, and M2T, respectively), (c,d,g,f) simulated spectra for different levels of water. The insets show a particular frame of the recorded evaporation process of a water droplet from Supplementary Movies M1R, M2R, M1T, and M2T. The white circles in the frames indicate the collection area of reflected/transmitted light in the experiment.}
  \label{fig:dynamicComp}
\end{figure*}
The sensitivity of the optical response related to the SLR to the thickness of the water layer on top of the metasurface can be used to achieve dynamic tuneability and real-time control of optical spectra and associated colouring of metasurfaces. 
%Although in our experimental setup is not possible to precisely control the level of water, it is possible to mimic such a process and test the dynamic optical response of the system. 
To demonstrate this, the evaporation of a water droplet deposited on top of the metasurface was monitored by recording the evolution of the transmittance and reflectance spectra every 100~ms until the droplet completely evaporated (Figs.~\ref{fig:dynamicComp}a,b,e,f). These experiments were conducted for the same metasurfaces described in the previous section, showing a gradual change of the spectral shape for both reflectance and transmittance. In particular, the time evolution of the reflectance spectra shows a gradual disappearance (metasurface M1, Fig.~\ref{fig:dynamicComp}a) and shift (metasurface M2, Fig.~\ref{fig:dynamicComp}b) of the lattice resonance as a function of time and related changes of the apparent colours. Simulated spectra for decreasing water levels show a similar trend as the dynamic evolution of the experimental spectra. (Figs.~\ref{fig:dynamicComp}c,d,g,h). The equivalent of Fig.~\ref{fig:dynamicComp} for the dynamic tunability of the metasurfaces M3 and M4 is reported in Fig. S11. Although the precise control of the water level was not investigated here, these results confirm that such control can provide dynamic tuning of metasurfaces' optical response.

Snapshots of the metasurfaces during the evaporation of water reveal that the evaporation process happens by forming a clear and curved front between air and water regions, showing a difference in colour (see Supplementary Movies) for real-time dynamic colour modification upon water evaporation. It is worth mentioning that in our experimental conditions (ambient temperature), water-to-air transition takes place over a few seconds (Supplementary Movies and Fig.~S12a,b of Supplementary Information). The process can be accelerated by controlling the temperature of the metasurface. The reflectance and transmittance simulations for water levels between 0 and 400~nm (Fig.~S12c,d of Supplementary Information) show that changes in the reflectance and transmittance spectral features are very sensitive to the water level (Fig.~S13 of Supplementary Information). These experiments, therefore, demonstrate the possibility of inferring water levels down to the nanometre scale based on the optical response of the metasurfaces. Furthermore, the uniformity of the colour across the metasurface confirms that the resonance condition does not involve long-range (far-field) contributions as already predicted in the theoretical derivations. This means that SLRs can develop in finite size metasurfaces, which may be beneficial for pixels in high-resolution displays.

\section*{Discussion}\label{concl}

We have proposed and demonstrated a practical and general strategy to dynamically control the optical response of an all-dielectric metasurface on a substrate by modifying its superstrate. In our case, this is done by varying the level of water on top of a metasurface, which makes our tuning approach completely reversible. In general, the SLR exists at the wavelength where the equality between the lattice sum and the inverse of the single particle polarizability is satisfied. In a homogeneous medium, the divergence of the lattice sum guarantees that this equality is always satisfied. However, when the metasurface is in an inhomogeneous medium, the lattice sum is finite and decreases with increasing the superstrate/substrate dielectric contrast due to the suppression of long-range (far-field) contributions on the lattice sum. This effectively decreases the number of nanoparticles participating in the SLR as the dielectric contrast increases, and opens possibilities for the realization and control of SLRs in finite size metasurfaces.

In the case of a metasurface with a static superstrate, the air and water cases considered in this paper open two options: (1) the SLR exists for both superstrates, and (2) the SLR exists only for one superstrate. In case (1), we demonstrated spectral shifting of the SLR, while in case (2) the SLR exhibits on/off switch. In both circumstances, the presence of the substrate plays a major role to achieve optical control.

In the case of a metasurface with a dynamic superstrate, the variation of the water level enables tunability of the optical response.
%Based on these theoretical results and full-wave simulations, several high-index silicon metasurfaces with different parameters were fabricated. The experimental measurements of these metasurfaces via progressive immersion in water and vice versa (drying) fully confirm our theoretical study. 
The experimental demonstration not only reveals drastic colour changes induced by the dynamic variation of the superstrate but also indicates the possibility of measuring water levels down to the nanometre scale by monitoring the optical response of the metasurface. Dielectric metasurfaces based on the tuning principles presented in this paper exhibit designed spectral selectivity and dynamic control of spectral response which may find application in colouring with small size pixels, high-resolution displays, sensors, wavelength-selective tuneable filters and mirrors, augmented reality and anticounterfeiting. 

\section*{Methods}

\subsection*{Analytical methods}

The behaviour of single particle resonances and individual multipole contributions for different excitation and environment conditions are discussed in Supplementary Information Section~1. A comprehensive theoretical description of the response of a metasurface in a homogeneous surrounding environment within the coupled-dipole method is presented in Supplementary Information Section~2. Supplementary Information Section~3 contains the detailed derivation of the lattice sum in presence of substrate in the far-field approximation (Eq.~\eqref{s_app}).

\subsection*{Numerical simulations}
The refractive index of air, water and glass used in the simulations are $n$=1, $n$=1.33 and $n$=1.45, respectively. The refractive index of Si can be found in Fig.~S1. 

Individual multipole contributions (multipole decomposition) of the single particle resonances are calculated in Ansys Lumerical using the expressions for the multipole moments presented in Ref.~\cite{Evlyukhin2019}. 

The scattering efficiency of a single particle, as well as the reflectance and transmittance of the array are obtained numerically using full-wave finite-difference time-domain (FDTD) simulations in Ansys Lumerical. For the single particle, a perfectly matched layer (PML) boundary condition is used for all boundaries. In the case of the array, periodic boundary conditions are used along the in-plane $x$- and $y$-directions. The fraction of reflectance and transmittance to each grating order is calculated by using the ``Grating order transmission" analysis group object of Ansys Lumerical.

Full dipole sum $\tilde{S}$ in presence of a substrate is numerically calculated as follows. First, we place an $x$-polarized dipole point source (with a default dipole moment $p_x$ set by the software, Ansys Lumerical) at ${\bf r}_0=(0,0,z_{\rm p})$ and apply periodic boundary conditions along the in-plane $x$ and $y$-directions with a periodicity $P$ (i.e., symmetrically respect to ${\bf r}_0$ at $x=y=\pm P/2$) and PML boundary condition in the $z$-direction. We then calculate the $x$-component of the total electric field $E_x^{\rm M}$ at ${\bf r}_0$. To exclude the central dipole ($l$=0) contribution, we repeat the above calculation with the PML boundary conditions in all Cartesian directions and subtract the electric field of the single dipole $E_x^{\rm 1}$ from the first simulation result $E_x^{\rm M}$. Finally, we calculate the full dipolar sum as $\tilde{S}=\varepsilon_0 \varepsilon_{\rm d} (E_x^{\rm M}-E_x^{\rm 1})/p_x$.

\subsection*{Experimental methods}

\subsubsection*{Sample design and fabrication}

The metasurfaces are fabricated by electron beam lithography on a 100 nm thick polycrystalline silicon film deposited on a glass substrate. Four different arrays of nanodisks with varying diameters and periods were considered (Table I). The square array size for each structure is 100~µm$\times$100~µm. 

%% Table of measured samples
%\begin{table}
%\begin{center}
%%\begin{minipage}{200pt}
%\caption{Geometrical parameters of the fabricated metasurfaces. Nanodisks height is $H$=100~nm for all metasurfaces.}\label{tab1}%
%\begin{tabular}{@{}llll@{}}
%\toprule
%Metasurface & $D$ (nm) & $P$ (nm) & Attribute\\
%\toprule
%M1    & 185   & 490  & Switching  \\
%M2    & 220   & 440  & Shift  \\
%M3    & 225   & 535  & Switching  \\
%M4    & 285   & 585  & Shift  \\
%\botrule
%\end{tabular}
%%\end{minipage}
%\end{center}
%\end{table}

\begin{table}
\caption{\label{tab:tab1} Geometrical parameters of the fabricated metasurfaces. Nanodisks height is $H$=100~nm for all metasurfaces.}
\begin{ruledtabular}
\begin{tabular}{llll}
Metasurface &$D$ (nm) &$P$ (nm) &Attribute\\
\hline
M1    & 185   & 490  & Switching  \\
M2    & 220   & 440  & Shift  \\
M3    & 225   & 535  & Switching  \\
M4    & 285   & 585  & Shift  \\
\end{tabular}
\end{ruledtabular}
\end{table}

\subsubsection*{Optical measurements}

Transmittance and reflectance measurements at normal incidence were performed by illuminating the sample using a tungsten-halogen white light source. The beam, polarised along the $x$-axis of the array of nanodisks, is weakly focused onto the sample using a microscope objective with numerical aperture 0.1. In the case of reflectance measurements, the reflected light is collected by the same objective while for transmittance measurements, the transmitted light is collected using a  microscope objective with numerical aperture 0.25. The collected light is then coupled to a spectrometer via an optical fibre. The size of the collection area is $\sim$75~µm$^{2}$. In both cases, a non-polarising beamsplitter is used to divert a fraction of the collected light onto a CCD camera to record images or movies of the sample (Fig.~S14). The spectral response of the nanostructures in water is measured by drop casting 50~µL of deionised water onto the nanostructure, fully immersing it. In order to capture the spectral response of the sample while the water level is changing in between the nanodisks, spectra are recorded every 100~ms until the water droplet has fully dried. Movies of the drying process of water on the nanostructures were simultaneously recorded.

%\backmatter

\section*{Supplementary information}

\noindent The paper is supported by the following additional materials:\\
Supplementary Information\\
Supplementary Movie-M1R\\
Supplementary Movie-M2R\\
Supplementary Movie-M3R\\
Supplementary Movie-M4R\\
Supplementary Movie-M1T\\
Supplementary Movie-M2T\\
Supplementary Movie-M3T\\
Supplementary Movie-M4T

\section*{Acknowledgments}
This work was supported, in part, by the Deutsche Forschungsgemeinschaft (DFG, German Research Foundation) under Germany’s Excellence Strategy within the Cluster of Excellence PhoenixD (EXC 2122, Project ID 390833453), Alexander von Humboldt Foundation, the ERC iCOMM project (789340) and the UK
EPSRC project EP/W017075/1. The authors acknowledge Cornerstone for the help in the fabrication of the samples.

%\section*{Declarations}
%Some journals require declarations to be submitted in a standardised format. Please check the Instructions for Authors of the journal to which you are submitting to see if you need to complete this section. If yes, your manuscript must contain the following sections under the heading `Declarations':

\section*{Funding}
Deutsche Forschungsgemeinschaft (DFG, German Research Foundation) under Germany’s Excellence Strategy within the Cluster of Excellence PhoenixD (EXC 2122, Project ID 390833453), Alexander von Humboldt Foundation, the ERC iCOMM project (789340) and the UK
EPSRC project EP/W017075/1.

% \section*{Competing interests}
% The authors declare no competing interests.

%\section*{Ethics approval}
%Not applicable

%\section*{Consent to participate}
%Not applicable

%\section*{Consent for publication}
%Not applicable

% \section*{Data availability}
% %%Not applicable
% All the data supporting this study are presented in the Results section and Supplementary Information and available from the corresponding authors upon reasonable request.

%\section*{Code availability}
%Not applicable

% \section*{Author contributions}
% I.A., A.E. and D.J.R. contributed equally to the work. I.A. performed numerical simulations, A.E. performed analytical studies, D.J.R. performed experiments, B.Ch., A.V.Z, A.E. and A.C.L. developed the idea. All authors contributed to writing the manuscript. 


\bibliography{References}
\end{document}

%%\unnumbered% uncomment this for unnumbered level heads
%Dynamic dielectric metasurfaces based on lattice resonances: tuning and switching effects via water
% Enabling optical tunability 
% Optical tunability and dynamics of Mie-type surface lattice resonances in a non-homogeneous environment.
\title{Supplementary Information \\ for \\ Dynamic dielectric metasurfaces via control of surface lattice resonances in non-homogeneous environment}

\author{Izzatjon Allayarov$^{\dag1,2,3}$, Andrey B. Evlyukhin$^{\dag3,4*}$, Diane J. Roth$^{\dag5}$, Boris Chichkov$^{3,4}$, Anatoly V. Zayats$^{5*}$ and Antonio Cal{\`a} Lesina$^{1,2,3*}$} 

\affiliation{$^1$Hannover Centre for Optical Technologies, Leibniz University Hannover, Nienburger Str. 17, D-30167 Hannover, Germany}
\affiliation{$^2$Institute of Transport and Automation Technology, Leibniz University Hannover, An der Universit{\"a}t 2, D-30823 Garbsen, Germany}
\affiliation{$^3$Cluster of Excellence PhoenixD, Leibniz University Hannover, Welfengarten 1A, D-30167 Hannover, Germany}
\affiliation{$^4$Institute of Quantum Optics, Leibniz University Hannover, Welfengarten 1, D-30167 Hannover, Germany}
\affiliation{$^5$Department of Physics and London Centre for Nanotechnology, King's College London, Strand, WC2R 2LS London, UK}

%\keywords{keyword1, Keyword2, Keyword3, Keyword4}

\maketitle
\def\thefootnote{*}\footnotetext{Corresponding author(s). E-mail(s): evlyukhin@iqo.uni-hannover.de, anatoly.zayats@kcl.ac.uk, antonio.calalesina@hot.uni-hannover.de}\def\thefootnote{\arabic{footnote}}
\def\thefootnote{$\dagger$}\footnotetext{These authors contributed equally to this work}\def\thefootnote{\arabic{footnote}}

\section{Single particle response}

\begin{figure}[b]
	\begin{center}
		\includegraphics[width=12cm]{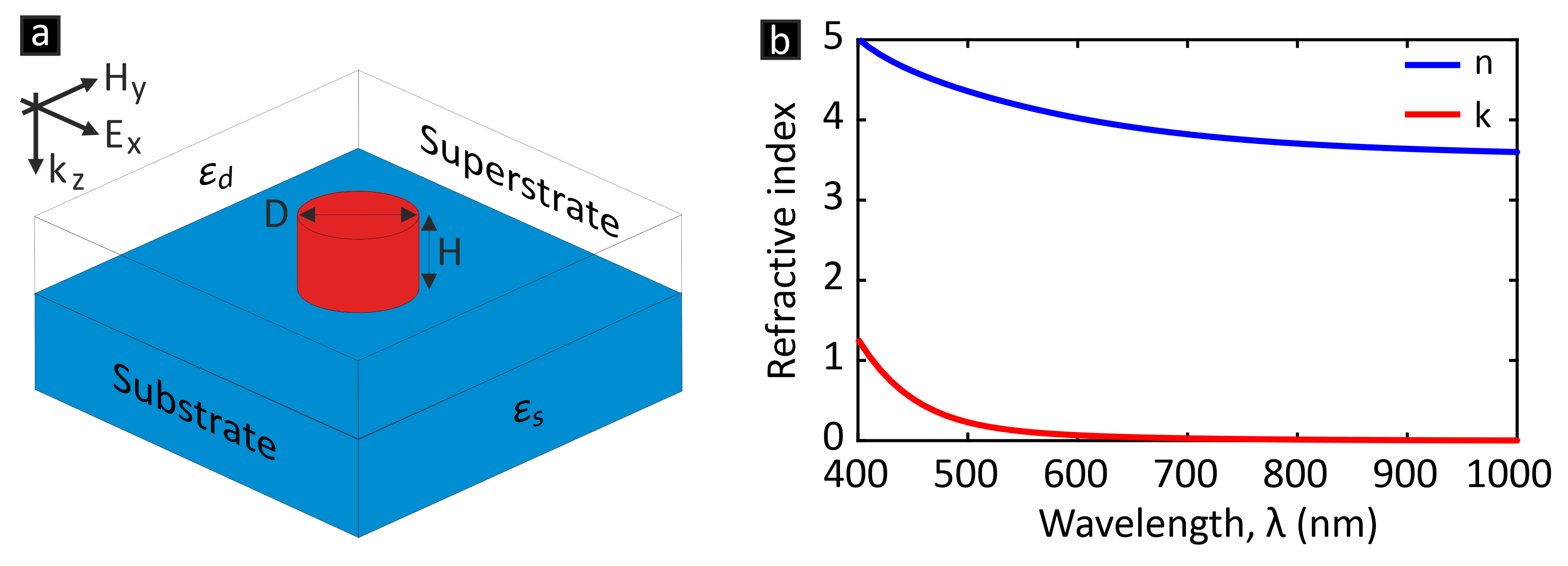}
		\caption{\textbf{Unit cell of a metasurface and refractive index of silicon.} (a) Sketch of the unit cell of a metasurface: Si nanodisk of diameter $D$ and height $H$ are placed on a glass substrate. The particle is excited at normal incidence, with $x$-linearly polarized light. (b) Real ($n$, blue) and imaginary ($k$, red) parts of the refractive index of polycrystalline silicon used in the simulations.}
		\label{fig:index}
	\end{center}
\end{figure}

In this section, we investigate the response of a single particle, which is used as the building block of the metasurface considered in the main text (Fig.~\ref{fig:index}a); polycrystalline Si is used in simulations (Fig.~\ref{fig:index}b). 

\begin{figure}[t]
	\begin{center}
		\includegraphics[width=12cm]{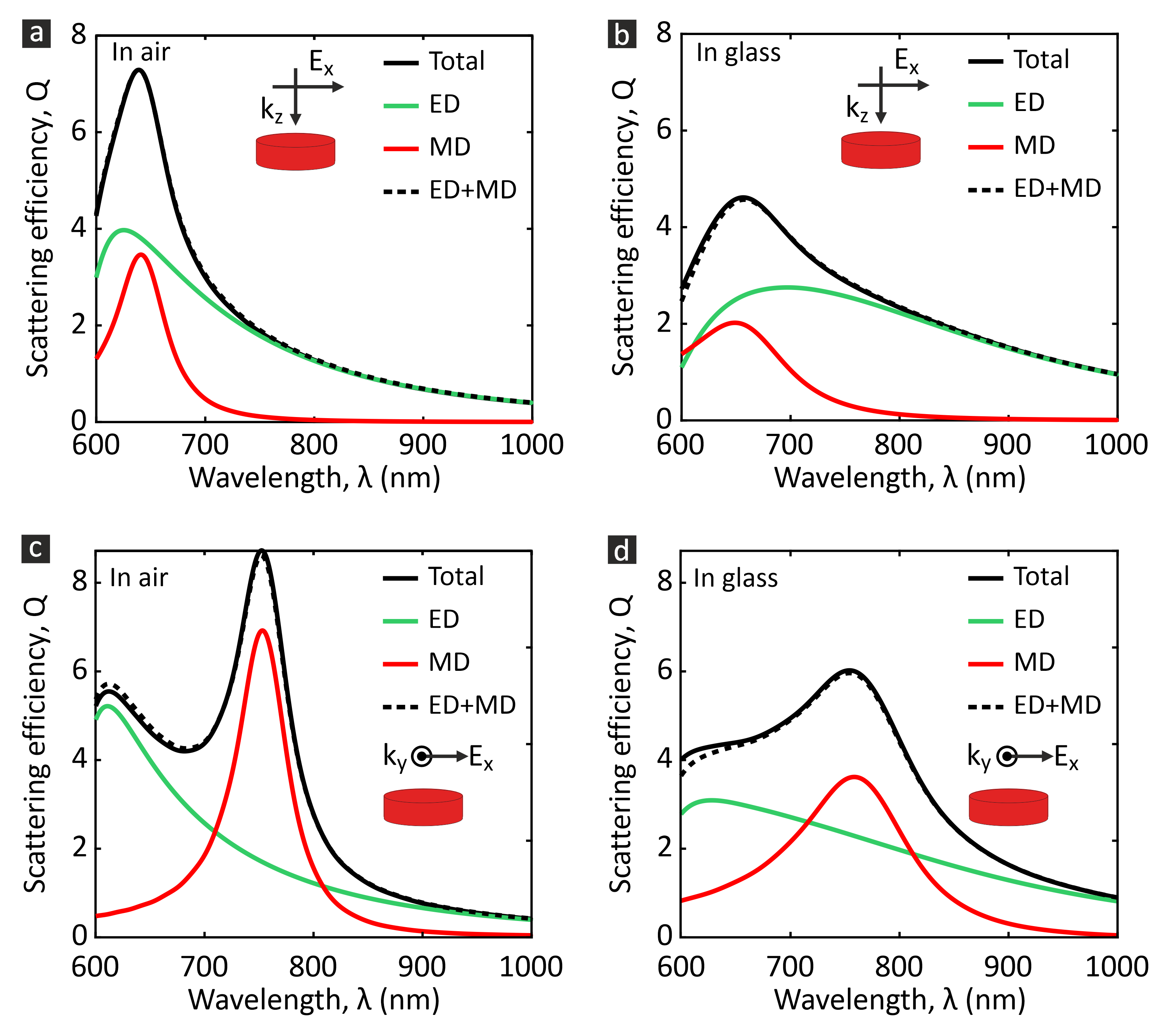}
		\caption{\textbf{Single particle resonances for different excitation configurations in a homogeneous medium.} Scattering efficiency ($Q=4\sigma D^{-2}/\pi$) spectrum of a single Si nanodisk with $D$=200~nm $H$=100~nm immersed in (a, c) air and (b, d) glass. The Si nanodisk is illuminated with a linearly polarized plane wave from (a, b) top and (c, d) side of the nanodisk. The scattering efficiency (black line) is decomposed into individual electric (ED, green line) and magnetic (MD, red line) dipole contributions.}
		\label{fig:sigma}
	\end{center}
\end{figure}

A Si nanodisk with $D$=100~nm in a homogeneous medium exhibits mainly electric and magnetic dipolar response for different (top, side) angles of incidence and in different surroundings (air, glass) relevant to this work (Fig.~\ref{fig:sigma}). 
%This allows us to consider the coupled-dipole method for the analytical description of the metasurface consisting of such nanodisks.

\begin{figure}[h]
	\begin{center}
		\includegraphics[width=12cm]{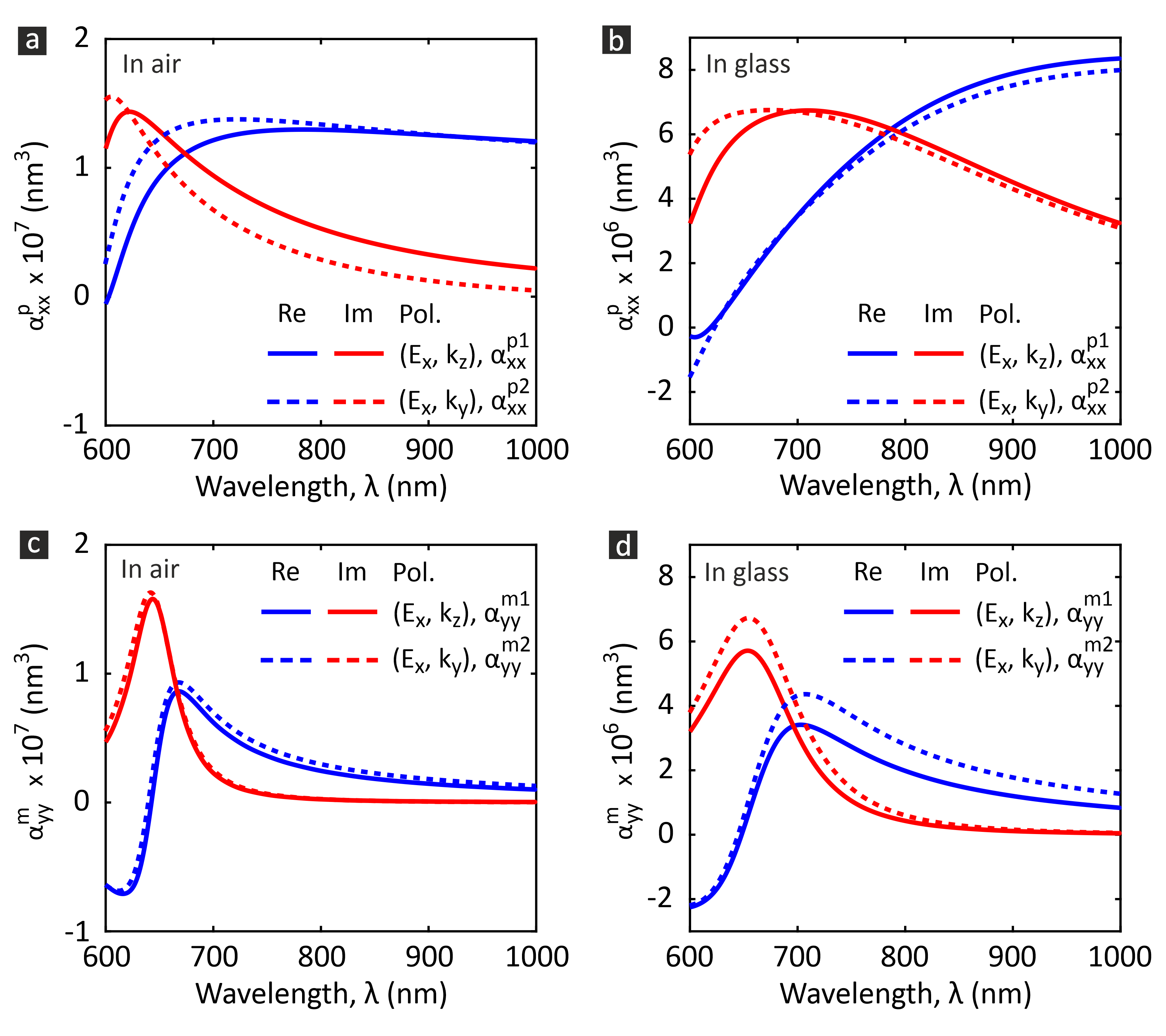}
		\caption{\textbf{Dipolar polarizabilities for different excitation conditions in a homogeneous environment.} Real (blue) and imaginary (red) parts of the electric dipole polarizabilities [panels (a) and (b)] of Si nanodisk with $D$=200~nm located in (a) air and (b) glass surrounding condition. Solid lines are for linearly-polarized plane wave excitation from the top ($E_x$ and $k_z$) of the nanodisk, while dashed lines are for the side ($E_x$ and $k_y$) excitation. Panels (c) and (d) show the same plots for the magnetic dipole polarizabilities in air and glass surrounding conditions, respectively.}
		\label{fig:alpha}
	\end{center}
\end{figure}

For the side and top excitation configurations, the corresponding dipole polarizabilities are similar (Fig.~\ref{fig:alpha}), therefore, the dependence of the dipolar polarizabilities on the excitation conditions, i.e, nonlocal effects, can be neglected~\cite{Bobylev2020}.

\begin{figure}[t]
	\begin{center}
		\includegraphics[width=12cm]{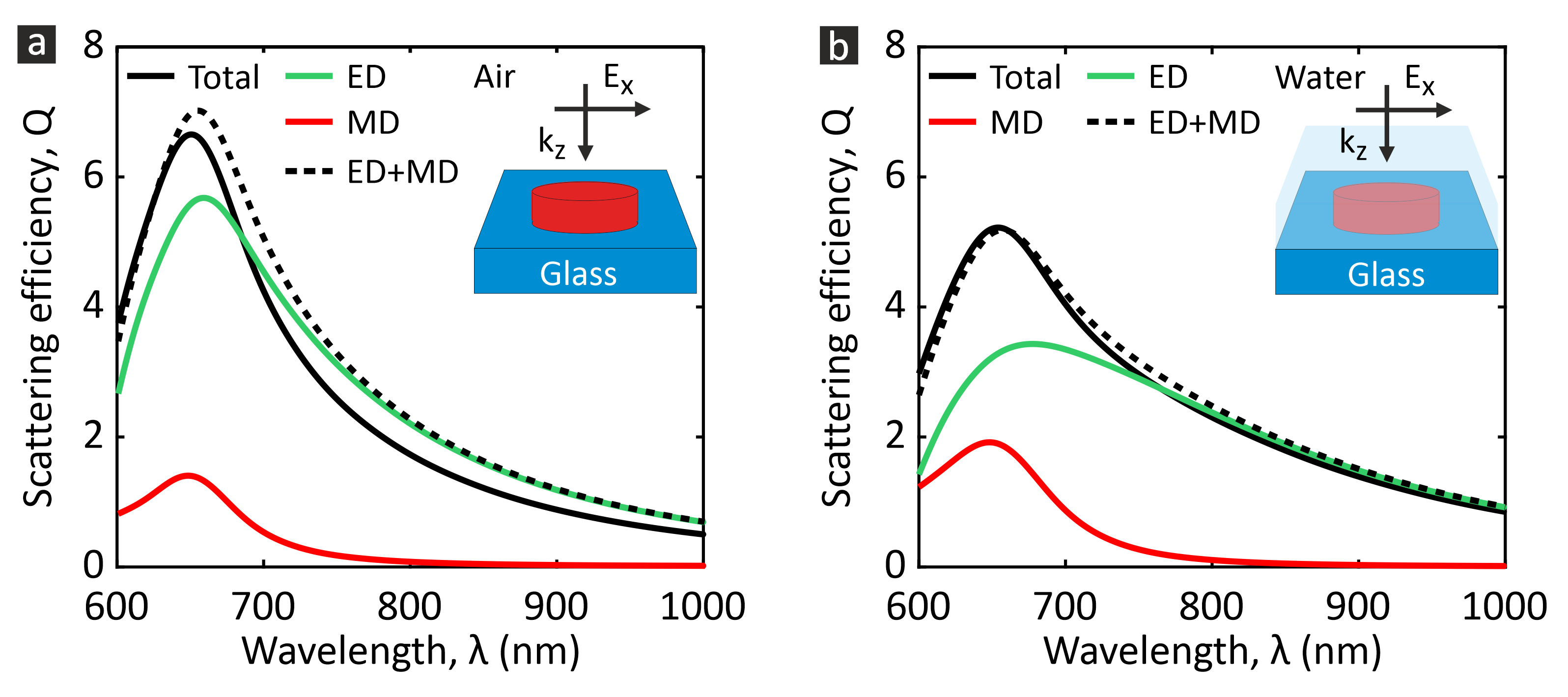}
		\caption{\textbf{Single particle resonances under normal incidence illumination and in an inhomogeneous environment.} Scattering efficiency ($Q=4\sigma D^{-2}/\pi$) spectrum of a single Si nanodisk with $D$=200~nm placed in (a) air and (b) water superstrate and on glass substrate. The Si nanodisk is irradiated with a linearly-polarized plane wave from the top of the nanodisk. The scattering efficiency (black line) is decomposed into individual electric (ED, green line) and magnetic (MD, red line) dipole contributions.}
		\label{fig:me200sub}
	\end{center}
\end{figure}

Similarly, the presence of a substrate does not significantly modify the nature of the response, which is again dominated by a dipolar response (Fig.~\ref{fig:me200sub}). Hence, the coupled-dipole method can be applied for the analytical description of the metasurface in a homogeneous environment and in the presence of a glass substrate. In Fig.~\ref{fig:me200sub}, only direct contribution of the electric and magnetic dipoles are considered, i.e., interference between multipoles occurring due to the reflection from the substrate is neglected~\cite{Pors2015}.
\begin{figure}[t]
	\begin{center}
		\includegraphics[width=12cm]{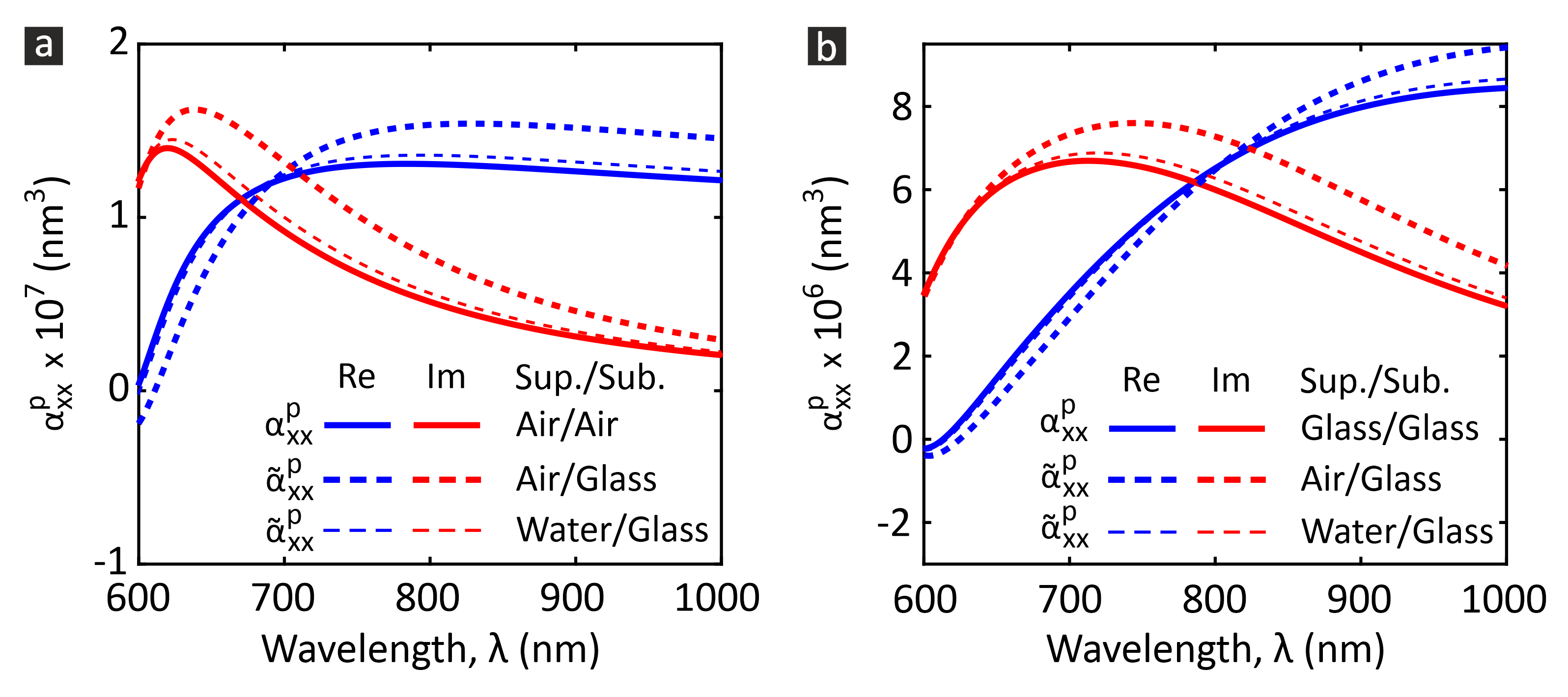}
		\caption{\textbf{Estimation of the electric dipole polarizability for different superstrate/substrate configurations.} Real (blue) and imaginary (red) parts of the electric dipole polarizabilities $\tilde\alpha_{xx}^{\rm p}$ (dashed lines) calculated based on Eq.~4 of the main text for air/glass (thin dashed lines) and water/glass (thick dashed lines) superstrate/substrate combinations. Solid lines correspond to the electric dipole polarizabilities $\alpha_{xx}^{\rm p}$ of Si nanodisk with $D$=200~nm located in (a) air and (b) glass homogeneous surrounding for linearly-polarized plane wave excitation from the top ($E_x$ and $k_z$) of the nanodisk.}
		\label{fig:alphaTilde}
	\end{center}
\end{figure}

Since the influence of a substrate on the dipole polarizabilities of Si nanodisks are weak (Fig.~\ref{fig:alphaTilde}), within the near-field approach, we can approximate a disk as an oblate ellipsoid with the polarizability $\tilde\alpha_{xx}^{\rm p}$ near the substrate as~\cite{Evlyukhin2005}
\begin{equation}
	\tilde\alpha_{xx}^{\rm p}\approx\left(\frac{1}{\alpha_{xx}^{\rm p}}+\frac{1}{4\pi(2z_{\rm p})^3}\frac{\varepsilon_{\rm d}-\varepsilon_{\rm s}}{\varepsilon_{\rm d}+\varepsilon_{\rm s}}\right)^{-1},
\end{equation}
where $z_{\rm p}$ is the distance between the substrate surface and the disk center, and $\alpha_{xx}^{\rm p}$ is the polarizability in a homogeneous environment with permittivity $\varepsilon_{\rm d}$, while $\varepsilon_{\rm s}$ is the permittivity of the substrate. As it is shown in Fig.~\ref{fig:alphaTilde}, the maximum relative difference between $\tilde\alpha_{xx}^{\rm p}$ (dashed lines) and $\alpha_{xx}^{\rm p}$ (solid lines) is $\leq$15\% [for the disk with $D$=200~nm and $H$=100~nm ($z_{\rm p}$=$H$/2)] in air superstrate and on the glass substrate (thick dashed lines). It becomes even smaller (less than 5\%) for the water superstrate (thin dashed lines). Hence, we can assume and indeed see later in the main text that the conditions of the lattice resonances for the arrays near the substrate are affected by the lattice sum $\tilde S$.

\clearpage
\section{Coupled-dipole model for a metasurface in a homogeneous environment}

For a homogeneous environment with a relative permittivity $\varepsilon_{\rm d}$ and illumination/polarisation conditions as in Fig.~\ref{fig:index}a, the equations for electric ${\bf p}=(p_x,0,0)$ and magnetic ${\bf m}=(0,m_y,0)$ dipole moments of a disk in a metasurface (with a square unit cell of period $P$) can be written as \cite{Abajo2007,Evlyukhin2010}
\begin{subequations}\label{ed}
	\begin{align}
		p_x&=\varepsilon_0\varepsilon_{\rm d}\alpha_{xx}^{\rm p1}E_x+\alpha_{xx}^{\rm p2}S_{xx}p_x,\label{ed:ed}\\
		m_y&=\alpha_{yy}^{\rm m1}H_y+\alpha_{yy}^{\rm m2}S_{yy}m_y,\label{ed:md}
	\end{align}
\end{subequations}
where $\varepsilon_0$ is the vacuum permittivity, $\alpha_{xx}^{\rm p1}$ ($\alpha_{xx}^{\rm p2}$) and $\alpha_{yy}^{\rm m1}$ ($\alpha_{yy}^{\rm m2}$) are the corresponding diagonal components of the electric and magnetic dipole polarizability tensors of a single disk for normal/top (lateral/side) incidence of external waves, respectively, $S_{xx}$ and $S_{yy}$ are the diagonal elements of the lattice dipole sum, $E_x$ and $H_y$ are the electric and magnetic fields of the incident wave at the geometric center of the disk located at the origin of the coordinate system. Note, here, we take into account that the components of the polarizability tensors can in general depend on the illumination direction. We consider a square elementary cell with $P_x=P_y=P$, therefore, $S_{xx}=S_{yy}=S$. From the solutions of Eq.~(\ref{ed})
\begin{subequations}\label{ed2}
	\begin{align}
		p_x&=\frac{\varepsilon_0\varepsilon_{\rm d} E_x}{1/\alpha_{xx}^{\rm p1}-\alpha_{xx}^{\rm p2}S/\alpha_{xx}^{\rm p1}},\label{ed2:ed}\\
		m_y&=\frac{H_y}{1/\alpha_{yy}^{\rm m1}-\alpha_{yy}^{\rm m2}S/\alpha_{yy}^{\rm m1}},\label{ed2:md}
	\end{align}
\end{subequations}
the spectral positions of the metasurface resonances are determined by 
% \begin{equation}\label{re}
	%     \mathrm{Re}(1/\alpha_{xx}^{\rm p1}-\alpha_{xx}^{\rm p2}S/\alpha_{xx}^{\rm p1})=0 \quad{\rm or}\quad \mathrm{Re}(1/\alpha_{yy}^{\rm m1}-\alpha_{yy}^{\rm m2}S/\alpha_{yy}^{\rm m1})=0. 
	% \end{equation}
\begin{subequations}\label{re}
	\begin{align}
		&\mathrm{Re}\Bigl(1/\alpha_{xx}^{\rm p1}-\alpha_{xx}^{\rm p2}S/\alpha_{xx}^{\rm p1}\Bigr)=\mathrm{Re}\Bigl(1/\alpha_{xx}^{\rm p,eff}\Bigr)=0, \label{re:ed}\\
		&\mathrm{Re}\Bigl(1/\alpha_{yy}^{\rm m1}-\alpha_{yy}^{\rm m2}S/\alpha_{yy}^{\rm m1}\Bigr)=\mathrm{Re}\Bigl(1/\alpha_{yy}^{\rm m,eff}\Bigr)=0, \label{re:md}
	\end{align}
\end{subequations}
where $\alpha_{xx}^{\rm p,eff}$ and $\alpha_{yy}^{\rm m,eff}$ are the effective electric and magnetic dipole polarizabilities, respectively. It should be noted that the imaginary part of the denominator of Eqs.~(\ref{ed2}) defines the quality factor of a resonance. If it is not sufficiently small, one may not see a distinct resonance even though the conditions Eqs.~(\ref{re}) are fulfilled.

Within the coupled-dipole approximation, the reflectance $R$ and transmittance $T$ of a metasurfaces can be written as~\cite{Evlyukhin2010}
\begin{equation}\label{RT}
	R=\left\lvert \frac{ik_{\rm d}}{2A} \Bigl( \alpha_{xx}^{\rm p,eff}-\alpha_{yy}^{\rm m,eff} \Bigr) \right\rvert ^2 \quad \mathrm{and} \quad T=\left\lvert 1-\frac{ik_{\rm d}}{2A}\Bigl( \alpha_{xx}^{\rm p,eff}+\alpha_{yy}^{\rm m,eff} \Bigr) \right\rvert ^2,
\end{equation}
% \begin{equation}\label{T}
	%     T=\left\lvert 1-\frac{ik_{\rm d}}{2A}\bigl( \alpha_{xx}^{\rm eff}+\alpha_{yy}^{\rm eff} \bigr) \right\rvert ^2,
	% \end{equation}
where $k_{\rm d}$ is the wave number in a homogeneous environment with $\varepsilon_{\rm d}$, $A$ is the unit cell area.

Disk polarizabilities $\alpha_{xx}^{\rm p}=p_x/(\varepsilon_0\varepsilon_{\rm d} E_x)$ and $\alpha_{yy}^{\rm m}=m_y/H_y$ can be determined numerically considering a single disk response (Fig.~\ref{fig:alpha}). Note, if the nonlocal effects of the disk response \cite{Bobylev2020} are negligible, $\alpha_{xx}^{\rm p1}=\alpha_{xx}^{\rm p2}$ and  $\alpha_{yy}^{\rm m1}=\alpha_{yy}^{\rm m2}$. Indeed, for the disks considered in Fig.~\ref{fig:sigma}, the electric and magnetic polarizabilities are similar for both top and side excitation conditions as in Fig.~\ref{fig:alpha}.

The lattice dipole sum $S$ for a homogeneous medium can be written as a sum of the individual far- $S^{\rm FF}$, middle- $S^{\rm MF}$ and near- $S^{\rm NF}$ field contributions \cite{Evlyukhin2012}:
% \begin{equation}\label{S}
	%    S=\frac{k_{\rm d}^2}{4\pi} \sum_{l\neq0}^\infty\frac{e^{ik_{\rm d}R_l}}{R_l}\left(1+\frac{i}{k_{\rm d}R_l}-\frac{1}{k_{\rm d} ^2 R_l^2}-\frac{x_l^2}{R_l^2}-\frac{3ix_l^2}{k_{\rm d}R_l^3}+\frac{3x_l^2}{k_{\rm d}^2R_l^4}\right),
	% \end{equation}
\begin{equation}\label{S}
	S=S^{\rm FF}+S^{\rm MF}+S^{\rm NF},
\end{equation}
with
\begin{subequations}\label{S1}
	\begin{align}
		S^{\rm FF}&=\frac{k_{\rm d}^2}{4\pi} \sum_{l\neq0}^\infty\frac{{\rm e}^{ik_{\rm d}R_l}}{R_l}\left(1-\frac{x_l^2}{R_l^2}\right),\label{S1:Sff}\\
		S^{\rm MF}&=\frac{k_{\rm d}^2}{4\pi} \sum_{l\neq0}^\infty\frac{{\rm e}^{ik_{\rm d}R_l}}{R_l^2}\left(\frac{i}{k_{\rm d}}-\frac{3ix_l^2}{k_{\rm d} R_l^2}\right),\label{S1:Smf}\\
		S^{\rm NF}&=\frac{k_{\rm d}^2}{4\pi} \sum_{l\neq0}^\infty\frac{{\rm e}^{ik_{\rm d}R_l}}{R_l^3}\left(-\frac{1}{k_{\rm d} ^2 }+\frac{3x_l^2}{k_{\rm d}^2R_l^2}\right),\label{S1:Snf}
	\end{align}
\end{subequations}
where the summation goes over all lattice nodes, excluding the node located at the origin of the coordinate system ($l=0$), $R_l$ is the distance between the node with number $l$ and the origin of the coordinate system, $x_l$ is the $x$-coordinate of the $l$-th node.
\begin{figure}[t]
	\begin{center}
		\includegraphics[width=12cm]{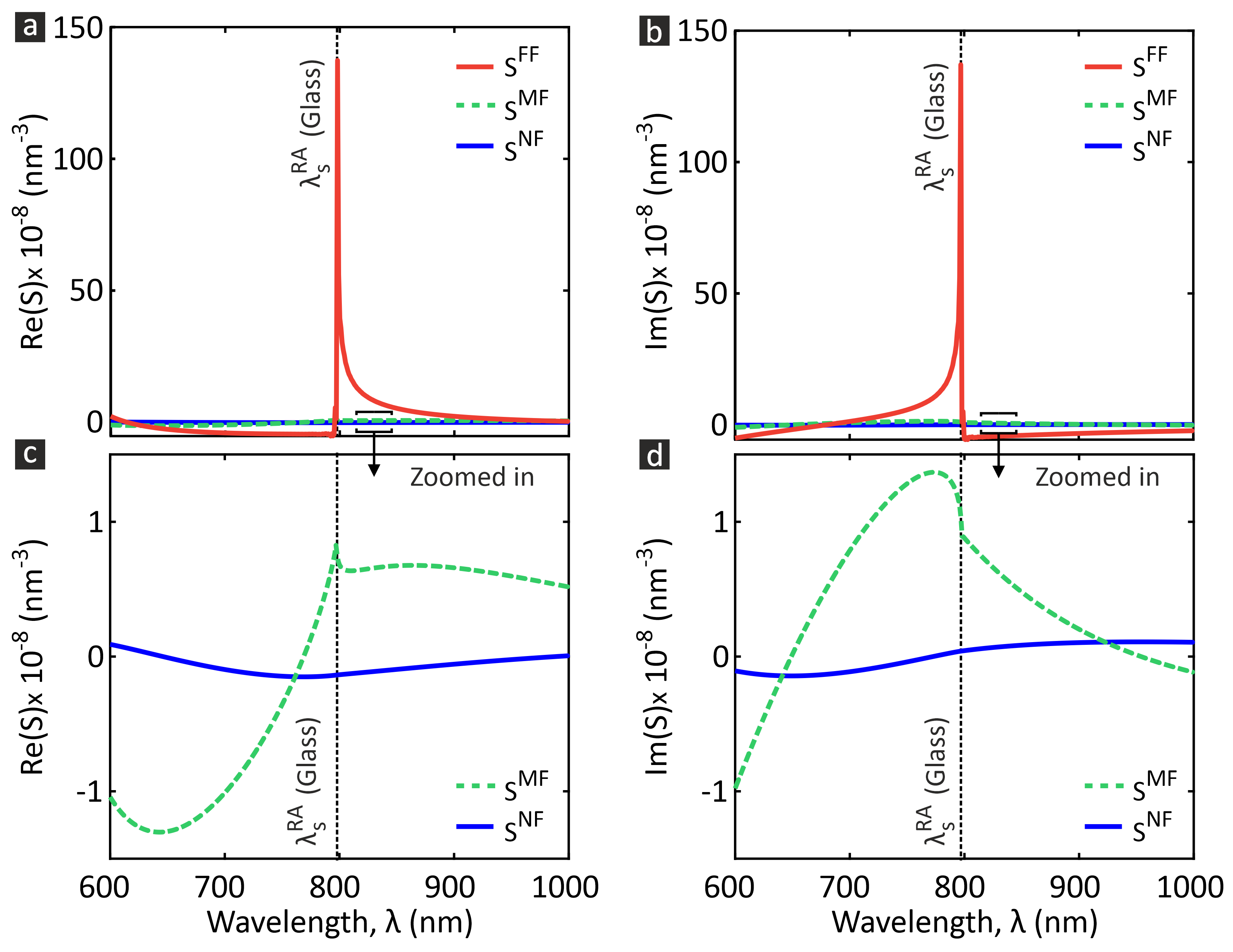}
		\caption{\textbf{Far-, middle- and near-field parts of the lattice dipole sum $S$.} (a) Real and (b) imaginary parts of the far- (red), middle- (green dashed) and near-field (blue) contributions of $S$ as a function of wavelength for the $x$-polarized dipoles located in glass. The vertical black dashed lines indicate the position of the Rayleigh anomaly $\lambda_{\rm d}^{\rm RA}$. Diameter and period of disks are $D$=200~nm and $P$=550~nm, respectively.}
		\label{fig:sum}
	\end{center}
\end{figure}
\begin{figure}[t]
	\begin{center}
		\includegraphics[width=12cm]{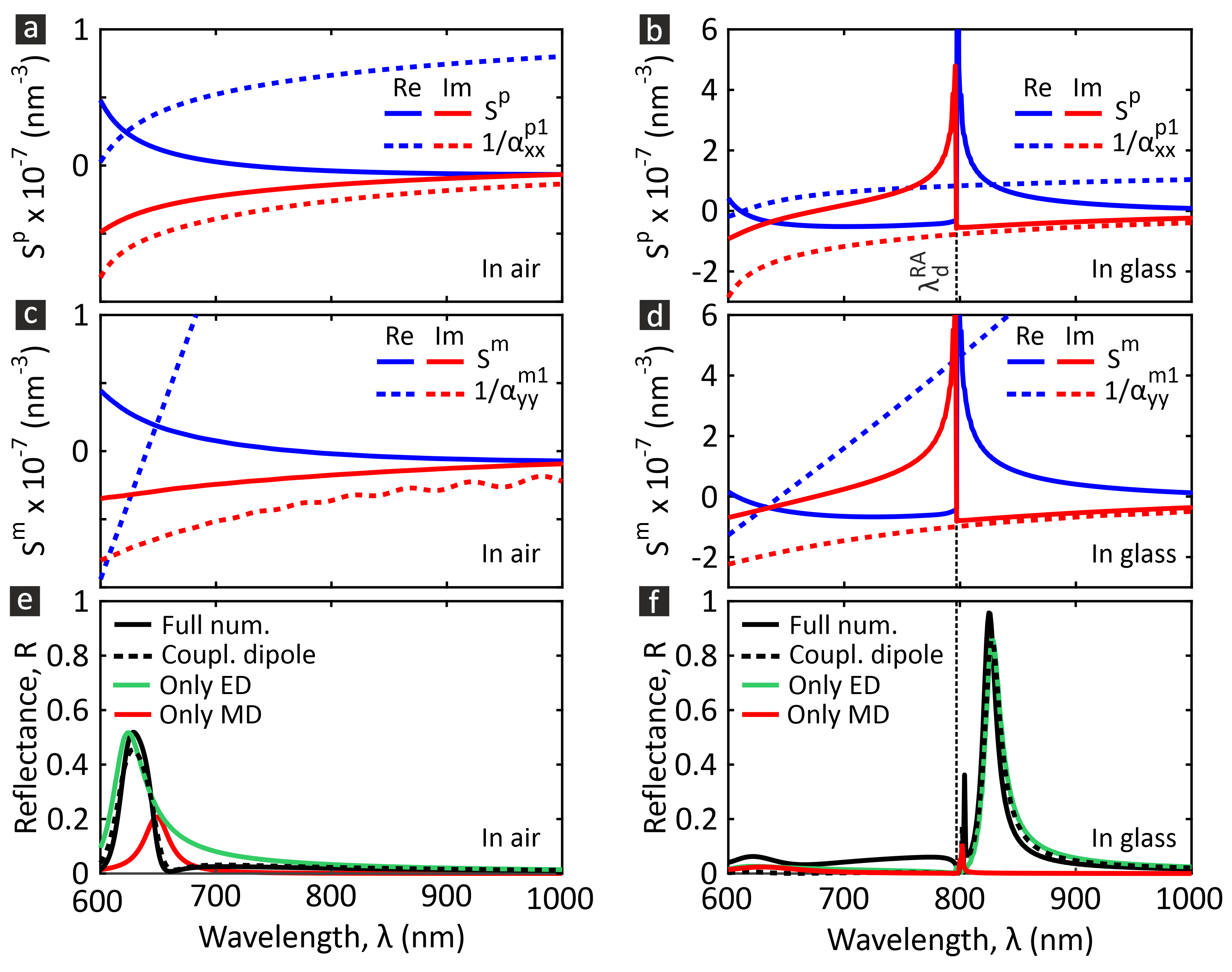}
		\caption{\textbf{Resonant conditions and spectra of the metasurface in homogeneous air and glass environments.} (a,b) Spectra of $1/\alpha_{xx}^{\rm p1}$ and $S^{\rm p}=\alpha_{xx}^{\rm p1} S/\alpha_{xx}^{\rm p2}$. (c,d) Spectra of $1/\alpha_{yy}^{\rm m1}$ and $S^{\rm m}=\alpha_{yy}^{\rm m1} S/\alpha_{yy}^{\rm m2}$. (e,f) The reflectance spectra of the metasurface calculated via the full-wave numerical simulations and the coupled dipole approximation with individual electric and magnetic dipole contributions. the metasurafce is in (a,c,e) air and (b,d,h) glass surroundings. The vertical black dashed line $\lambda_{\rm d}^{\rm RA}$ indicates the spectral position of the Rayleigh anomaly in glass. The disk parameters are $D$=200~nm, $H$=100~nm and $P$=550~nm, respectively.}
		\label{fig:me}
	\end{center}
\end{figure}

By comparing these three contributions with the total lattice sum (Fig.~\ref{fig:sum}) for the $x$-polarized dipoles located in glass with a period of $P$=550~nm, one can identify that only the far-field contribution (red line) diverges at the Rayleigh anomaly wavelength ($\lambda^{\rm RA}=Pn_{\rm d}$, where $n_{\rm d}$ is the refractive index of surrounding medium) in contrast to the middle- (green dashed line) and near- (blue line) field parts. The divergent nature of $S^{\rm FF}$ ensures that Eq.~(\ref{re}) has a solution.

Let us apply the model to predict and calculate the resonances of an infinite array of Si nanodisks with $D$=200~nm and $P$=550~nm placed in air or glass homogeneous surroundings (Fig.~\ref{fig:me}). In the former case, for the considered wavelength range, the diffractionless regime is realised. Figures~\ref{fig:me}a and \ref{fig:me}c visualize the roots of Eqs.~(\ref{re}a-b), which define the spectral position of the electric and magnetic lattice resonances, respectively. The metasurface's predicted resonances (the crossing point of blue lines) are around 625~nm for the electric dipole and around 650~nm for the magnetic dipole. Indeed, from the reflectance calculations in Fig.~\ref{fig:me}(e), we clearly see the individual electric (green) and magnetic (red) dipole resonances at those positions. The intensity of the magnetic dipole resonance is lower than the electric one. This is because for the magnetic dipole, the difference between the imaginary parts (red lines) of the dipole sum and inverse polarizability is larger than the electric dipole case. Furthermore, the total reflectance within the coupled-dipole approximation [Eq.~(\ref{RT}), black dashed line] has an excellent agreement with the full numerical (black solid line) calculation.

In the case of a glass environment, there are three resonant conditions in Fig.~\ref{fig:me}b, d: one at the Rayleigh anomaly position $\lambda_{\rm d}^{\rm RA}$ (vertical dashed line) and one on each sides of it. Because of the diffraction scattering into the homogeneous surrounding, resonances are not excited or strongly damped (large differences in the imaginary parts) in $\lambda \leq \lambda_{\rm d}^{\rm RA}$ region. In $\lambda > \lambda_{\rm d}^{\rm RA}$ region, we see sharp electric and magnetic resonances (Fig.~\ref{fig:me}f) due to the minimal difference in the imaginary parts. Note that the magnetic resonance (at $\sim$800~nm) quality factor is higher than the electric one (at $\sim$825~nm). This is because its spectral position is closer to $\lambda_{\rm d}^{\rm RA}$. However, as in the case of air surrounding, the magnetic dipole resonances are much weaker than the electric ones. Very good agreement between the coupled-dipole model and the full numerical simulations is observed.

\section{Derivation of $\tilde S$ in the far-field approximation}

The lattice sum, which is essentially a Green's function, of a layered system can be semi-analytically calculated in the Fourier space~\cite{Paulus2000,Novotny2012,mahi2017depth}. Here, it is convenient to work in the real space. The lattice sum $\tilde S$ in presence of a substrate can be presented as a sum of homogeneous $S$ [see Eq.~\eqref{S}] and additional sum $S^{\rm R}$ accounting for the reflection from the substrate. In the far-field approximation, $S^{\rm R}$ can be calculated analytically if one considers the electric far-field radiated along the surface of a flat substrate by an electric dipole moment $p_x$ placed near this surface at the point ${\bf r}_0$. Using Eqs.~(B1) and (B2) as well as Eqs.~(B5) and (B6) from Ref.~\cite{Evlyukhin2013}, we obtain the $x$-component of the electric far field $E_x$ at the point with spherical coordinates $(r,\theta,\varphi)$ radiated by $p_x$ placed at ${\bf r}_0=(0,0,z_{\rm p})$, as a sum of the electric far field $E_x^0$ of the wave directly propagating in the superstrate
\begin{equation}\label{ef}
	E_x^0(r,\theta,\varphi) = \frac{k_0^2}{4\pi\varepsilon_0 r}{\rm e}^{ik_{\rm d} r}{\rm e}^{-ik_{\rm d}z_{\rm p}\cos\theta} p_x (\sin^2\varphi+\cos^2\theta \cos^2\varphi), 
\end{equation}
and the electric far field $E_x^{\rm R}$ of the wave reflected from the substrate
\begin{equation}\label{es}
	E_x^{\rm R}(r,\theta,\varphi) = \frac{k_0^2}{4\pi\varepsilon_0 r}{\rm e}^{ik_{\rm d} r} {\rm e}^{ik_{\rm d}z_{\rm p}\cos\theta}  p_x (r^{({\rm s})} \sin^2\varphi - r^{({\rm p})}\cos^2\theta \cos^2\varphi),
\end{equation}
with the reflection coefficients for s- and p-polarized waves
\begin{subequations}\label{rcoef}
	\begin{gather}
		r^{\rm (s)}=\frac{\cos\theta-\sqrt{\varepsilon_{\rm s}/\varepsilon_{\rm d}-\sin^2\theta}}{\cos\theta+\sqrt{\varepsilon_{\rm s}/\varepsilon_{\rm d}-\sin^2\theta}},\label{rcoef:rs}\\
		r^{\rm (p)}=\frac{\varepsilon_{\rm s}\cos\theta-\varepsilon_{\rm d}\sqrt{\varepsilon_{\rm s}/\varepsilon_{\rm d}-\sin^2\theta}}{\varepsilon_{\rm s}\cos\theta+\varepsilon_{\rm d}\sqrt{\varepsilon_{\rm s}/\varepsilon_{\rm d}-\sin^2\theta}},\label{rcoef:rp}
	\end{gather}
\end{subequations}
where one considers that $r>>\lvert{\bf r}_0\rvert$, and $k_0$ and $k_{\rm d}$ are the vacuum and superstrate wave numbers, respectively, and  ${\bf n}={\bf r}/r=(\sin\theta \cos\varphi,\sin\theta \sin\varphi,\cos\theta)$ is the unit vector directed to the observation point.

From the reciprocity principle, one can show that the electric field at the point ${\bf r}$ created by the dipole point source located at ${\bf r}_0$ is equivalent to the electric field at ${\bf r}_0$ created by the same dipole located at ${\bf r}$, i.e., ${\bf p}({\bf r}_0,\omega) \cdot {\bf E}({\bf r},\omega)={\bf p}({\bf r},\omega) \cdot {\bf E}({\bf r}_0,\omega)$. Using Eqs.~(\ref{ef}) and (\ref{es}), we can obtain the analytical estimation of the far-field part $\tilde  S^{\rm FF}$ of $\tilde  S = S+S^{\rm R}$. Considering 
$\tilde S^{\rm FF} = S^{\rm FF}+S^{\rm R,FF}$, we have
\begin{align}\label{s_ff1}
	S^{\rm FF} &= \frac{\varepsilon_0 \varepsilon_{\rm d}}{p_x}\sum_{l\neq0}^\infty E^0_x({\bf r}_l) \\ \notag
	&= \frac{k_{\rm d}^2}{4\pi}\sum_{l\neq0}^\infty \frac{{\rm e}^{ik_{\rm d} R_l}}{R_l} {\rm e}^{-ik_{\rm d} z_{\rm p}^2/R_l} \left [1 -\frac{x_l^2}{R_l^2}\left (1 -\frac{z_{\rm p}^2}{R_l^2}\right) \right ],
\end{align}

\begin{align}\label{s_ff2}
	S^{\rm R,FF} &= \frac{\varepsilon_0 \varepsilon_{\rm d}}{p_x}\sum_{l\neq0}^\infty E^{\rm Ref}_x({\bf r}_l)\\ \notag
	&= \frac{k_{\rm d}^2}{4\pi}\sum_{l\neq0}^\infty \frac{{\rm e}^{ik_{\rm d} R_l}}{R_l} {\rm e}^{ik_{\rm d} z_{\rm p}^2/R_l}\left [r^{({\rm s})}_l-\frac{x_l^2}{R_l^2}\left (r^{({\rm s})}_l+r^{({\rm p})}_l\frac{z_{\rm p}^2}{R_l^2}\right) \right ],
\end{align}
in Cartesian coordinates and the reflection coefficients for s- and p-polarized waves generated by the dipoles located at points  are\cite{Evlyukhin2013}
\begin{subequations}\label{rl}
	\begin{gather}
		r^{({\rm s})}_l=\frac{z_{\rm p}/R_l-\sqrt{\Delta\varepsilon/\varepsilon_{\rm d}+z_{\rm p}^2/R_l^2}}{z_{\rm p}/R_l+\sqrt{\Delta\varepsilon/\varepsilon_{\rm d}+z_{\rm p}^2/R_l^2}},\label{rl:rs} \\ 
		r^{({\rm p})}_l=\frac{\varepsilon_{\rm s}z_{\rm p}/R_l-\varepsilon_{\rm d}\sqrt{\Delta\varepsilon/\varepsilon_{\rm d}+z_{\rm p}^2/R_l^2}}{\varepsilon_{\rm s}z_{\rm p}/R_l+\varepsilon_{\rm d}\sqrt{\Delta\varepsilon/\varepsilon_{\rm d}+z_{\rm p}^2/R_l^2}},\label{rl:rp}
	\end{gather}
\end{subequations}
where the electric fields $E^0_x({\bf r}_l)$ and $E^{\rm Ref}_x({\bf r}_l)$ are determined by Eqs.~(\ref{ef}) and (\ref{es}), respectively, with ${\bf r}_l=(x_l,y_l,z_{\rm p})$, $R_l=\lvert{\bf r}_l \rvert$, $\cos\theta_l=z_{\rm p}/R_l$ and $\Delta\varepsilon=(\varepsilon_{\rm s}-\varepsilon_{\rm d})$.

\begin{figure}[b]
	\begin{center}
		\includegraphics[width=12cm]{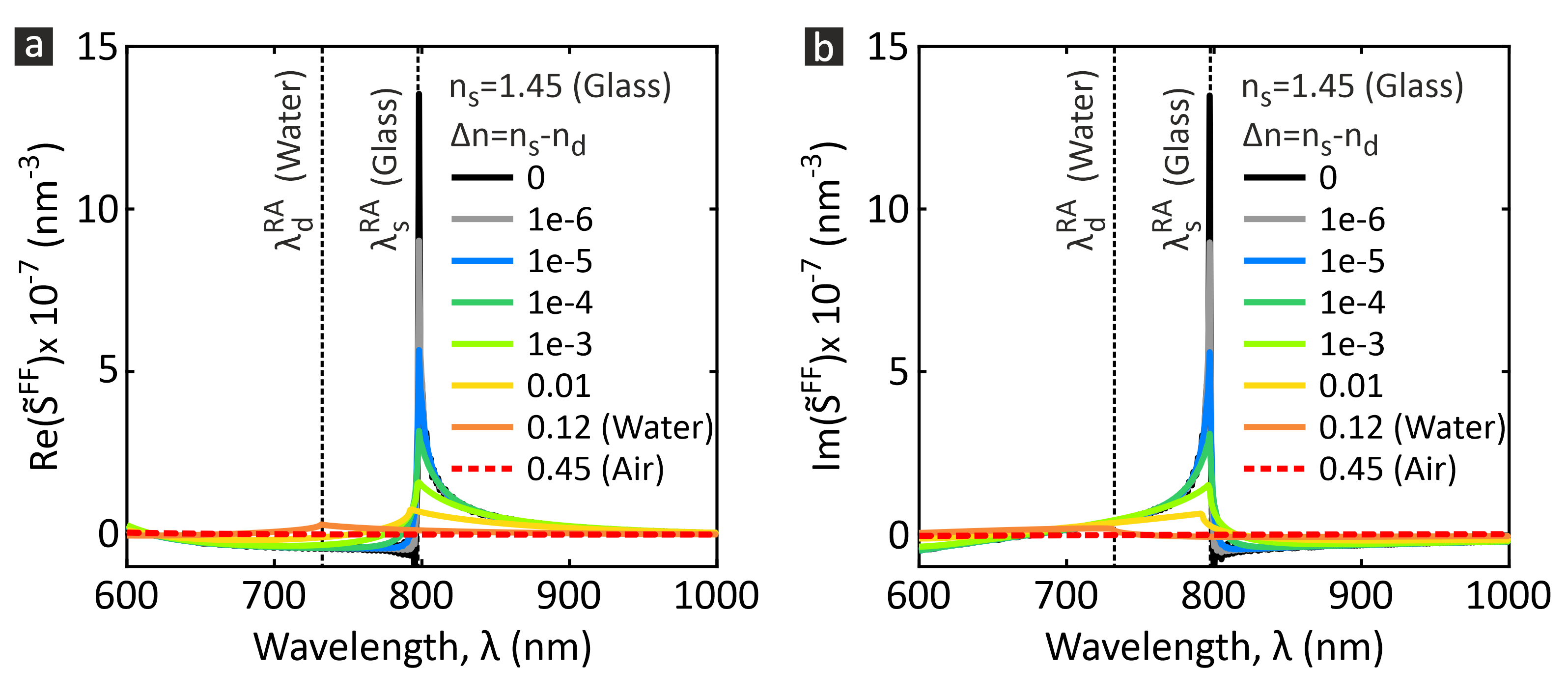}
		\caption{\textbf{Far-field lattice sum for different substrate/superstrate dielectric contrasts.} Spectra of (a) Real and (b) imaginary parts of the far-field lattice sum $\tilde S^{\rm FF}$ in presence of a substrate for a different refractive index contrast between superstrate ($n_{\rm d}$) and substrate ($n_{\rm s}$), $\Delta n=(n_{\rm s}-n_{\rm d})$. The vertical black dashed lines indicate the Rayleigh anomaly wavelengths in a glass substrate and a water superstrate.}
		\label{fig:sff}
	\end{center}
\end{figure}

As we can see from the spectra of the far-field lattice dipole sum $\tilde S^{\rm FF}$ for a different superstrate refractive indices $n_{\rm d}$ and a glass substrate with $n_{\rm s}$=1.45 (Fig.~\ref{fig:sff}), $\tilde S^{\rm FF}$ diverges at the Rayleigh anomaly $\lambda^{\rm RA}_{\rm s}$ at $n_{\rm d}=n_{\rm s}$ (red line) since $r^{({\rm s})}=r^{({\rm p})}=0$ and $\tilde S^{\rm FF}=S^{\rm FF}$ (see also Fig.~\ref{fig:sum}), while it converges at $n_{\rm d} \neq n_{\rm s}$.

Under the condition $R_l>>z_{\rm p}$, i.e., the period $P$ of the metasurface is much larger than the position of the particle centers relative to the substrate surface, Eqs.~(\ref{s_ff1}) and (\ref{s_ff2}) can be written as
\begin{equation}\label{s_ff1_approx}
	S^{\rm FF} \approx \frac{k_{\rm d}^2}{4\pi}\sum_{l\neq0}^\infty \frac{e^{ik_{\rm d} R_l}}{R_l} e^{-ik_{\rm d} z_{\rm p}^2/R_l} \!\left [1 -\frac{x_l^2}{R_l^2} \right ],
\end{equation}
\begin{equation}\label{s_ff2_approx}
	S^{\rm R,FF} \approx \frac{k_{\rm d}^2}{4\pi}\sum_{l\neq0}^\infty \frac{e^{ik_{\rm d} R_l}}{R_l}  e^{ik_{\rm d} z_{\rm p}^2/R_l}\left [1-\frac{x_l^2}{R_l^2} \right ]r^{({\rm s})}_l.
\end{equation}
% Under these conditions, the contributions of the reflection coefficients $r_l^{({\rm p})}$ for p-polarized waves are proportional to $\cos^2\theta_l=(z_{\rm p}/R_l)^2$ and are negligibly small. If also $k_{\rm d}z_{\rm p}\leq 1$ we can write
% \begin{equation}\label{s_app}
	%     \tilde S^{\rm FF}=S^{\rm FF}+S^{\rm Ref,FF}\approx  \frac{k_{\rm d}^2}{4\pi}\sum_{l\neq0}^\infty \frac{e^{ik_{\rm d} R_l}}{R_l}  \!\left [1 \!-\!\frac{x_l^2}{R_l^2}\! \right ] \Bigl (1+r_l^{({\rm s})}\Bigr).
	% \end{equation}
%and the reflection coefficients for s- and p-polarized waves
%\begin{equation}\label{rsl}
%   r^{({\rm s})}_l=\frac{z_{\rm p}-\sqrt{R_l^2(\varepsilon_{\rm s}/\varepsilon_{\rm d}-1)+z_{\rm p}^2}}{z_{\rm p}+\sqrt{R_l^2(\varepsilon_{\rm s}/\varepsilon_{\rm d}-1)+z_{\rm p}^2}}\:,
%\end{equation}
%\begin{equation}\label{rpl}
%    r^{({\rm p})}_l=\frac{z_{\rm p} \varepsilon_{\rm s} -\varepsilon_{\rm d}\sqrt{R_l^2(\varepsilon_{\rm s}/\varepsilon_{\rm d}-1)+z_{\rm p}^2}}{z_{\rm p} \varepsilon_{\rm s} +\varepsilon_{\rm d}\sqrt{R_l^2(\varepsilon_{\rm s}/\varepsilon_{\rm d}-1)+z_{\rm p}^2}}\:,
%\end{equation}
%where the electric fields $E^0_x({\bf r}_l)$ and $E^R_x({\bf r}_l)$ are determined by (\ref{ef}) and (\ref{es}), respectively, with ${\bf r}_l=(x_l,y_l,z_{\rm p})$, and $R_l=|{\bf r}_l|$.
%The above equations are written in the Cartesian coordinates using ${\bf r}_0\!=\!(0,0,z_{\rm p})$, $\cos\varphi=x_l/R_l$ and $\cos\theta=z_{\rm p}/R_l$.

In order to assess the role of the substrate in Eq.~\eqref{s_ff2_approx}, we introduce an effective distance $\Tilde{R}=z_{\rm p}\sqrt{{\varepsilon_{\rm d}}/{\lvert\Delta\varepsilon\rvert}}$, that  relates the position of the disk centers with respect to the substrate surface ($z_{\rm p}$) and the dielectric contrast $\Delta\varepsilon$, and consider the following limiting cases for $r_l^{({\rm s})}$, taking into account that $\varepsilon_{\rm s}$ and $\varepsilon_{\rm d}$ have only real values in the considered systems:\\
(1) $\Delta\varepsilon=0$, $\tilde{R}=\infty\quad\Rightarrow\quad r_l^{({\rm s})}=0$.\\
(2) $R_l=\tilde{R}$: \\
\indent 2.1) $\Delta\varepsilon<0\quad\Rightarrow\quad r_l^{({\rm s})}=1$,\\
\indent 2.2) $\Delta\varepsilon>0\quad\Rightarrow\quad r_l^{({\rm s})}=(1-\sqrt{2})/(1+\sqrt{2}) \approx -0.2$;\\
\noindent
(3) $R_l<<\tilde{R}$, $\Delta\varepsilon\neq0\quad\Rightarrow\quad \lvert r_l^{({\rm s})}\rvert<<1$;\\
\noindent
(4) $R_l>>\tilde{R}$, $\Delta\varepsilon\neq0\quad\Rightarrow\quad r_l^{({\rm s})}\simeq-1$;
\noindent
% 4) $\Delta\varepsilon=0\quad\Rightarrow\quad r_l^{({\rm s})}=0$ for any $l$ (the case of homogeneous environment).\\

% \noindent Using these limit cases, we can estimate the dipole sum $\tilde S$ for metasurfaces near a substrate. Importantly, for infinite metasurfaces placed  near a substarte  the condition 3) will always be satisfied starting from some large $R_l$ for which $r_l^{({\rm s})}$ is practically equal to $-1$. This means that the number of terms in $\tilde S^{\rm FF}$ can be considered finite and the total lattice sum $\tilde S$ will not diverge. In order to estimate the value of $\tilde S^{\rm FF}$, first, the parameter $\tilde R$ must be calculated, then it should be compared  with  the metasurface period $P$. If it turns out that $\tilde R<<P$ then one can consider that $\tilde S^{\rm FF}$ is negligible. If vice-versa $\tilde R>>P$ then  for an estimation of $\tilde S^{\rm FF}$ one can take into account the terms with $R_l\leq \tilde R$. \textcolor{red}{IA: We need to say something about other cases.}

In the case (1)--a metasurface in a homogeneous environment--there is no reflection and $S^{\rm R,FF}=0$. As was considered in the previous section, $S^{\rm FF}$ diverges at the position of the RA, and a solution of Eq.~(\ref{re}) exists, so that a surface-lattice resonance (SLR) can be excited in the spectral  region $\lambda>\lambda^{\rm RA}$ independently on a single particle polarizability \cite{markel2005divergence}. Furthermore, a divergent nature of the far field-field part is conditional upon the contribution of the dipoles at $\tilde{R}\Rightarrow\infty$. From the practical point of view, this means that metasurface has to be sufficiently large in order to guarantee the SLR excitation.

% \noindent The cases 2)-4) directly concern the substrate presence and can be used for estimation of the reflection coefficient $r_l^{({\rm s})}$ and $\tilde S^{\rm FF}$.

The case (2) is an intermediate situation where the reflection coefficient has a certain fixed value and it does not depend on ${\lvert\Delta\varepsilon\rvert}$.

The case (3) corresponds to the short-distance terms of the lattice sum. Since $\rvert r_l^{({\rm s})}\rvert<<1$, the influence of the substrate on these terms  is negligible.

The case (4)--a metasurface on a substrate--will always be realized  starting from some large $R_l$ for which $r_l^{({\rm s})}$ is  practically equal to $-1$ and $\tilde S^{\rm FF}_l\simeq 0$. 
This means that the contribution of the terms with large $R_l$ is negligible and $\tilde S^{\rm FF}$ has to be finite for the total lattice sum $\tilde S$ not to diverge. Hence, the far-field interparticle interaction is negligible and near- and middle-field interactions dominate. Thus, the value of $\tilde S$ and SLR can be controlled by a choice of $\Delta\varepsilon$. This also means that a metasurface can be relatively compact, which makes such systems even more attractive.

\begin{figure}[htbp]
	\begin{center}
		\includegraphics[width=12cm]{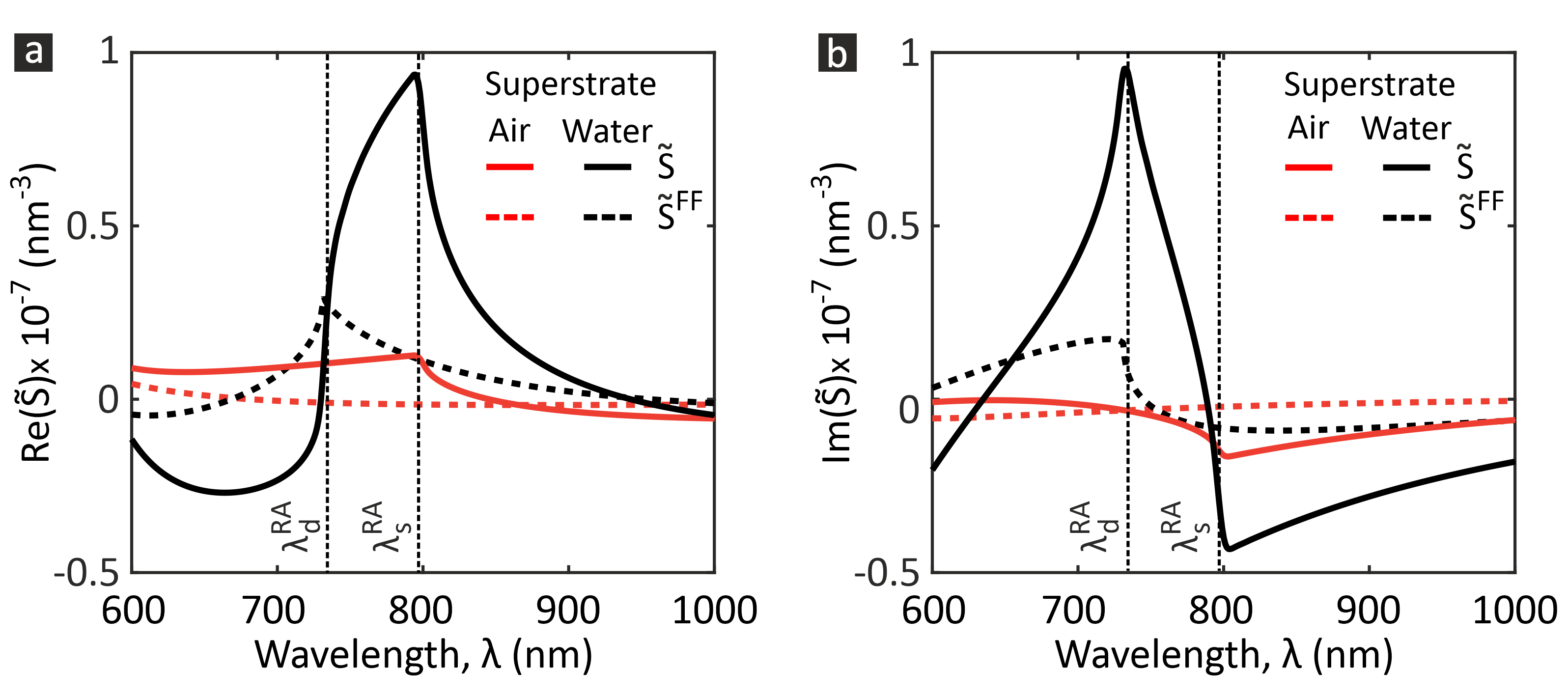}
		\caption{\textbf{Dipole lattice sum and resonance condition analysis in presence of a substrate.} Spectra of (a) real and (b) imaginary parts of the numerically calculated full lattice sum $\tilde S$ (solid lines) and its far-field part $\tilde S^{\rm FF}$ (dashed lines) for the air (red) and water (black) superstrate. The lattice period is $P$=550~nm.}
		\label{fig:s_substrate}
	\end{center}
\end{figure}

Notice that $\tilde S$ (which is calculated numerically, see Methods) is higher than $\tilde S^{\rm FF}$ (which is calculated analytically via Eq.~3) due to the non-negligible contribution of near- and medium-field terms (Fig.~~\ref{fig:s_substrate}).

\clearpage
\section{Additional experimental and simulation results}

\begin{figure}[htbp]
	\begin{center}
		\includegraphics[width=12cm]{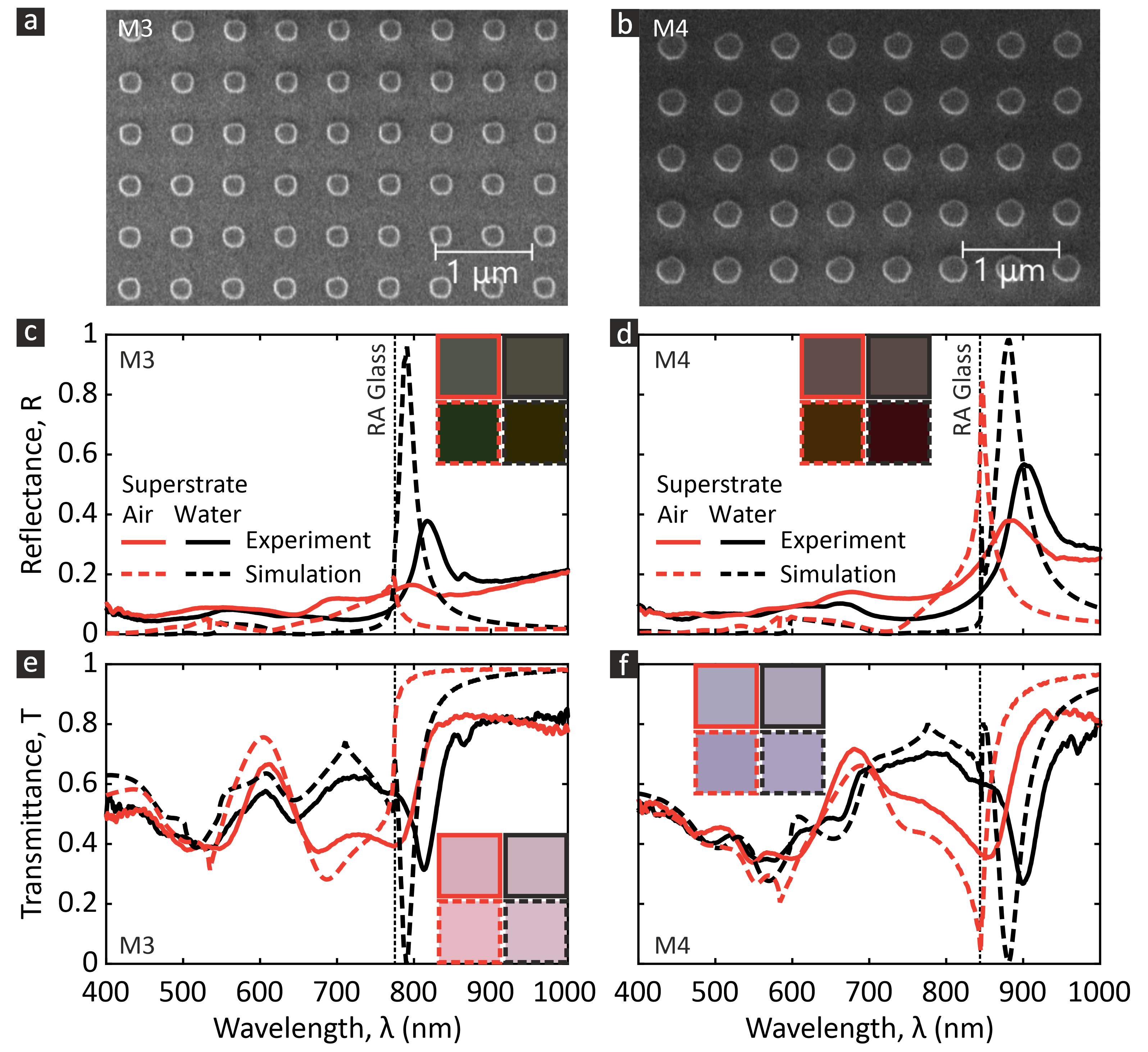}
		\caption{\textbf{Transmission and reflection of the metasurface with different superstrates.} (a,b) Scanning electron microscope (SEM) images of the fabricated metasurfaces: (a) M3 with $D$=225~nm and $P$=535~nm and (b) M4 with $D$=285~nm and $P$=585~nm. (c,d) Reflectance and (e,f) transmittance of metasurfaces (c,e) M3 and (d,f) M4 for (red curves) air and (black curves) water superstrates: (solid lines) experiment, (dashed lines) simulations. The vertical dashed lines  corresponds to the spectral position of the RAs of a glass substrate $\lambda_{\rm s}^{\rm RA}$. Insets shows the color representation of the corresponding spectra. In the simulations, an infinite array is considered and only the main diffraction order contribution is taken into account, while the measurement results correspond to the finite size, 100~µm$\times$100~µm metasurface.}
		\label{fig:steadyComp2}
	\end{center}
\end{figure}

\begin{figure}[htbp]
	\begin{center}
		\includegraphics[width=12cm]{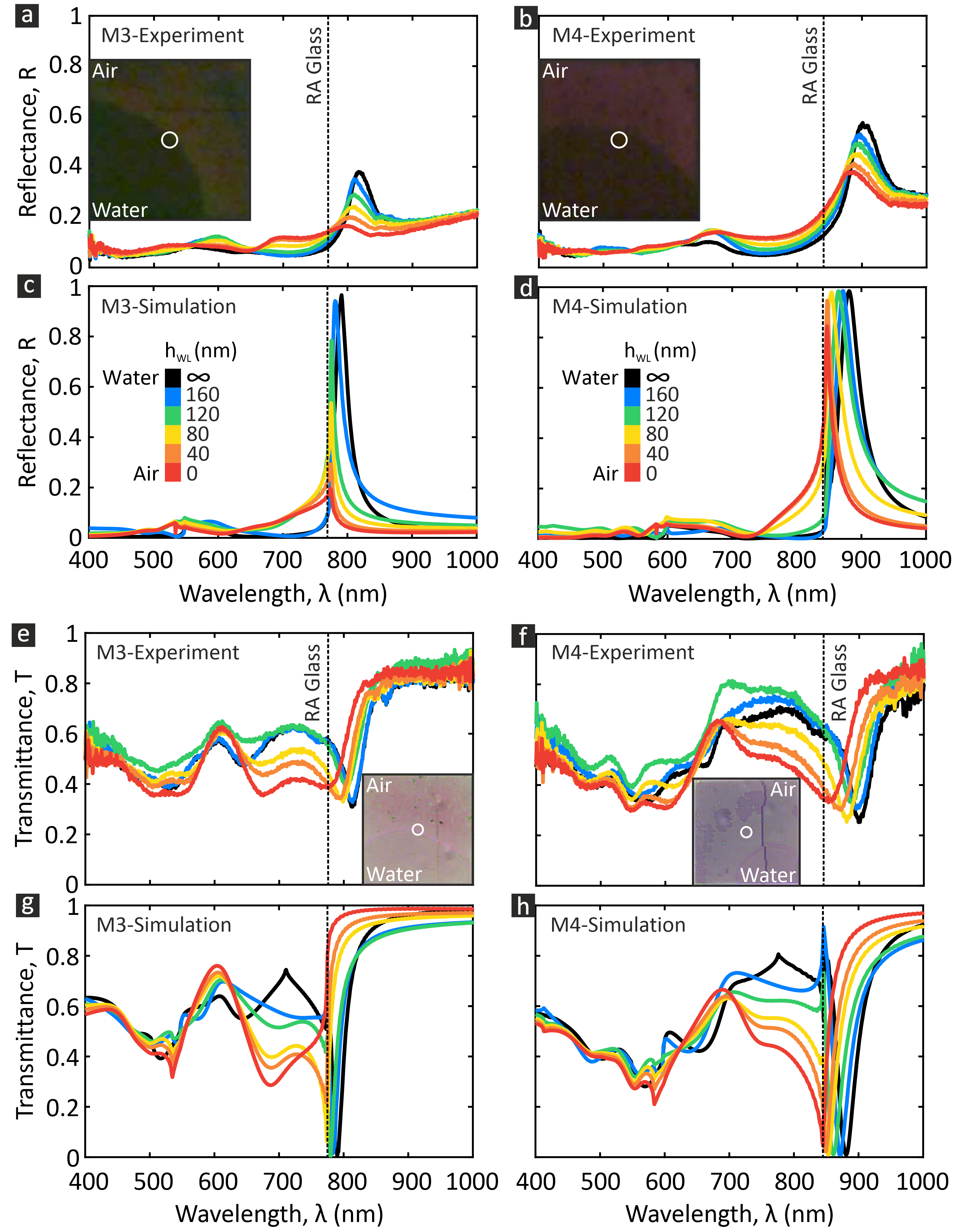}
		\caption{\textbf{Dynamic tuning of the transmission and reflection of the metasurface via superstrate control.} (a--d) Reflectance and (e--h) transmittance of metasurfaces (a,c,e,g) M3 and (b,d,f,h) M4 while dynamically changing a superstrate between air and water: (a,b,e,f) experimental spectra measured every 100~ms (snapshot of the Supplementary Movies M3R, M4R, M3T, and M4T, respectively), (c,d,g,f) simulated spectra for different levels of water. The insets show a particular frame of the recorded evaporation process of a water droplet from Supplementary Movies M3R, M4R, M3T, and M4T. The white circles in the frames indicate the collection area of reflected/transmitted light in the experiment. In the simulations, an infinite array is considered and only the main diffraction order contribution is taken into account, while the measurement results correspond to the finite size, 100~µm$\times$100~µm metasurface.}
		\label{fig:dynamicComp2}
	\end{center}
\end{figure}

\begin{figure}[htbp]
	\begin{center}
		\includegraphics[width=12cm]{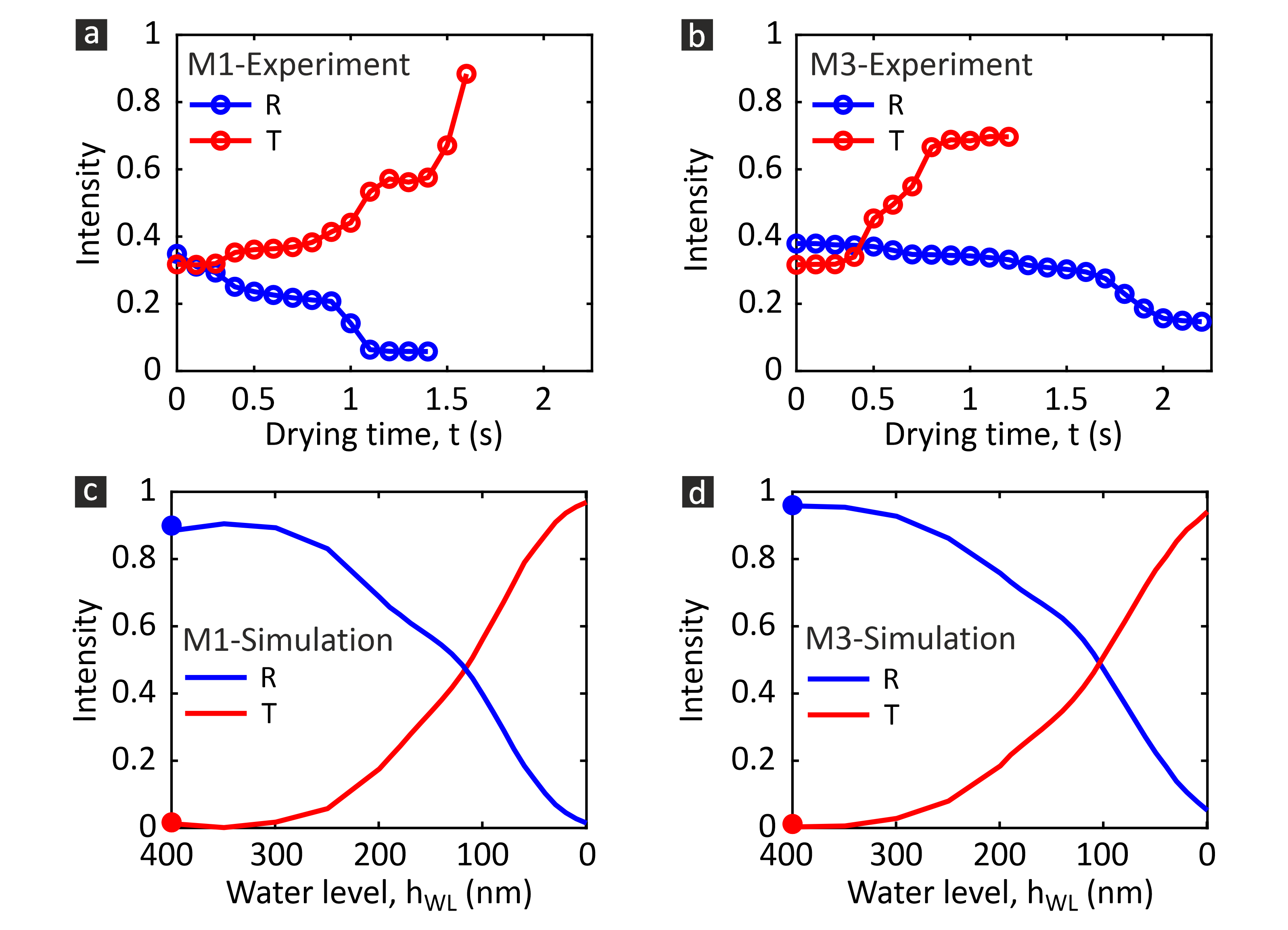}
		\caption{\textbf{Resonant response of a metasurface as a function of water level and drying time.} The reflectance ($R$, blue lines) and transmittance ($T$, red lines) at the resonance wavelength $\lambda$ as a function of the water level (top panels, simulation) and drying time (bottom panels, experiment) for (a, c) metasurface M1 ($\lambda$=735~nm and (b, d) metasurface M3 ($\lambda$=815~nm. The solid circles in (a, b) indicate the corresponding reflectance and transmittance at the resonance wavelength for an infinite water level.}
		\label{fig:rtWL}
	\end{center}
\end{figure}

\begin{figure}[htbp]
	\begin{center}
		\includegraphics[width=12cm]{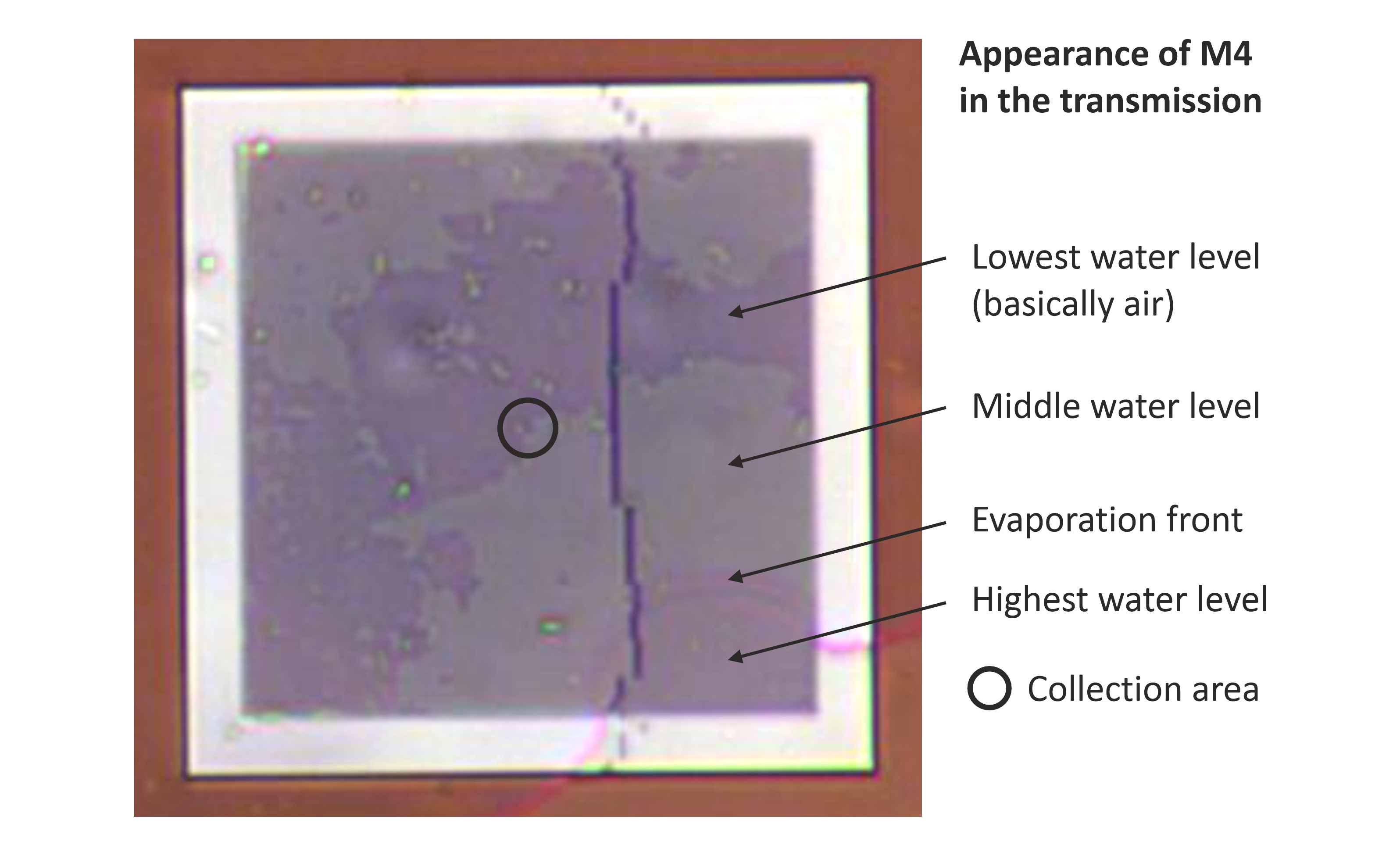}
		\caption{\textbf{Snapshot of the appearance of M4 in the transmission.} The snapshot (Supplementary Movie-M4T) shows the metasurface response for the different levels of water.}
		\label{fig:snapM4T}
	\end{center}
\end{figure}

\begin{figure}[htbp]
	\begin{center}
		\includegraphics[width=12cm]{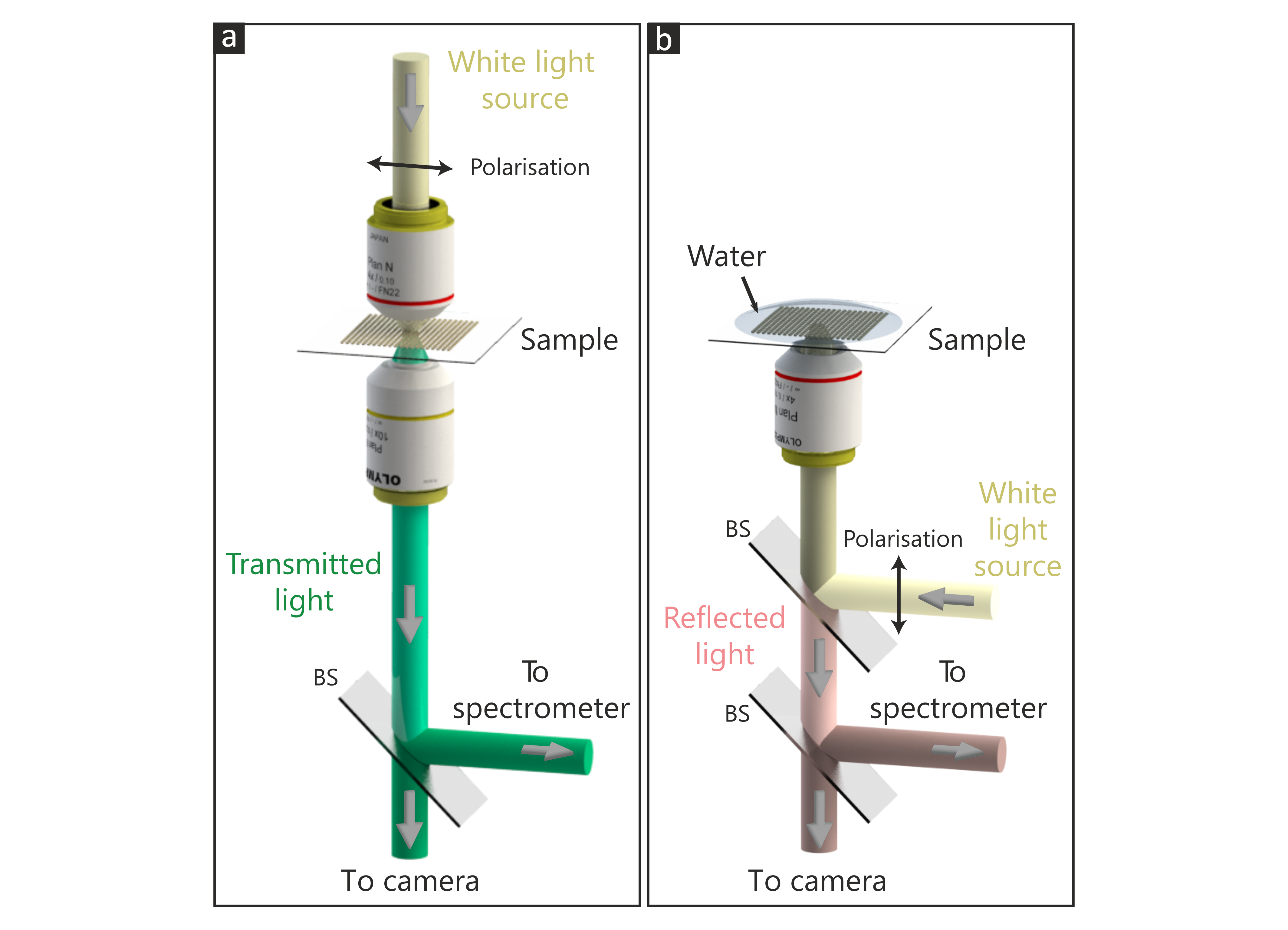}
		\caption{\textbf{Schematics of the experimental setup.} Diagram of the experimental setup for (a) transmittance and (b) reflectance measurements (BS: beamsplitter).}
		\label{fig:expSetup}
	\end{center}
\end{figure}


\clearpage
\bibliography{References}% common bib file